\documentclass[aps, reprint, superscriptaddress, amsmath, amssymb, floatfix, a4paper, hidelinks]{revtex4-1}
\usepackage[utf8]{inputenc}
\usepackage{amsfonts}
\usepackage{graphicx}
\usepackage{ifthen}
\usepackage{subcaption}
\usepackage{hyperref}
\usepackage{siunitx}
\usepackage[left=1.50cm, right=1.50cm, top=1.50cm, bottom=1.50cm]{geometry}
\usepackage[dvipsnames]{xcolor}
\graphicspath{{Figures/}}

\newcommand{\diff}[1]{\operatorname{d}\ifthenelse{\equal{#1}{}}{\,}{\!#1}}

\begin{document}
	\title{Johnson-Nyquist Noise Effects in Neutron Electric-Dipole-Moment Experiments}	
	
	\author{N.~J.~Ayres}
	\affiliation{ETH Z\"{u}rich, Institute for Particle Physics and Astrophysics, CH-8093 Z\"{u}rich, Switzerland}
	
	\author{G.~Ban}	
	\affiliation{Normandie Universit\'e, ENSICAEN, UNICAEN, CNRS/IN2P3, LPC Caen, 14000 Caen, France}
	
	\author{G.~Bison}
	\affiliation{Paul Scherrer Institut, CH-5232 Villigen PSI, Switzerland}
	
	\author{K.~Bodek}
	\affiliation{Marian Smoluchowski Institute of Physics, Jagiellonian University, 30-348 Cracow, Poland}
	
	\author{V.~Bondar}
	\affiliation{ETH Z\"{u}rich, Institute for Particle Physics and Astrophysics, CH-8093 Z\"{u}rich, Switzerland}
	
	\author{P.-J.~Chiu}
	\email[Corresponding author: ]{pin-jung.chiu@psi.ch}
	\affiliation{ETH Z\"{u}rich, Institute for Particle Physics and Astrophysics, CH-8093 Z\"{u}rich, Switzerland}
	\affiliation{Paul Scherrer Institut, CH-5232 Villigen PSI, Switzerland}	
	
	\author{B.~Clement}
	\affiliation{Universit\'e Grenoble Alpes, CNRS, Grenoble INP, LPSC-IN2P3, 38026 Grenoble, France}
	
	\author{C.~B.~Crawford}
	\affiliation{Department of Physics and Astronomy, University of Kentucky, Lexington, Kentucky, 40506, USA}
	
	\author{M.~Daum}
	\affiliation{Paul Scherrer Institut, CH-5232 Villigen PSI, Switzerland}
	
	\author{S.~Emmenegger}
	\affiliation{ETH Z\"{u}rich, Institute for Particle Physics and Astrophysics, CH-8093 Z\"{u}rich, Switzerland}
	
	\author{M.~Fertl}
	\affiliation{Institute of Physics, Johannes Gutenberg University Mainz, 55128 Mainz, Germany}
	
	\author{A.~Fratangelo}
	\affiliation{Laboratory for High Energy Physics and Albert Einstein Center for Fundamental Physics, University of Bern, CH-3012 Bern, Switzerland}
	
	\author{W.~C.~Griffith}
	\affiliation{Department of Physics and Astronomy, University of Sussex, Falmer, Brighton BN1 9QH, United Kingdom}
	
	\author{Z.~D.~Gruji\'{c}}
	\affiliation{Institute of Physics Belgrade, University of Belgrade, 11080 Belgrade, Serbia}
	
	\author{P.~G.~Harris}
	\affiliation{Department of Physics and Astronomy, University of Sussex, Falmer, Brighton BN1 9QH, United Kingdom}
	
	\author{K.~Kirch}
	\affiliation{ETH Z\"{u}rich, Institute for Particle Physics and Astrophysics, CH-8093 Z\"{u}rich, Switzerland}
	\affiliation{Paul Scherrer Institut, CH-5232 Villigen PSI, Switzerland}	
	
	\author{P.~A.~Koss}
	\altaffiliation[Present address: ]{Fraunhofer-Institut f\"{u}r Physikalische Messtechnik IPM, 79110 Freiburg i. Breisgau, Germany}
	\affiliation{Instituut voor Kern- en Stralingsfysica, University of Leuven, B-3001 Leuven, Belgium}
	
	\author{B.~Lauss}
	\affiliation{Paul Scherrer Institut, CH-5232 Villigen PSI, Switzerland}
	
	\author{T.~Lefort}	
	\affiliation{Normandie Universit\'e, ENSICAEN, UNICAEN, CNRS/IN2P3, LPC Caen, 14000 Caen, France}
	
	\author{P.~Mohanmurthy}
	\altaffiliation[Present address: ]{University of Chicago, 5801 S Ellis Ave, Chicago, IL 60637, USA.}
	\affiliation{ETH Z\"{u}rich, Institute for Particle Physics and Astrophysics, CH-8093 Z\"{u}rich, Switzerland}
	\affiliation{Paul Scherrer Institut, CH-5232 Villigen PSI, Switzerland}	
	
	\author{O.~Naviliat-Cuncic}
	\affiliation{Normandie Universit\'e, ENSICAEN, UNICAEN, CNRS/IN2P3, LPC Caen, 14000 Caen, France}
	
	\author{D.~Pais}
	\affiliation{ETH Z\"{u}rich, Institute for Particle Physics and Astrophysics, CH-8093 Z\"{u}rich, Switzerland}
	\affiliation{Paul Scherrer Institut, CH-5232 Villigen PSI, Switzerland}	
	
	\author{F.~M.~Piegsa}
	\affiliation{Laboratory for High Energy Physics and Albert Einstein Center for Fundamental Physics, University of Bern, CH-3012 Bern, Switzerland}
	
	\author{G.~Pignol}
	\affiliation{Universit\'e Grenoble Alpes, CNRS, Grenoble INP, LPSC-IN2P3, 38026 Grenoble, France}
	
	\author{D.~Rebreyend}
	\affiliation{Universit\'e Grenoble Alpes, CNRS, Grenoble INP, LPSC-IN2P3, 38026 Grenoble, France}
	
	\author{I.~Rien\"{a}cker}
	\affiliation{ETH Z\"{u}rich, Institute for Particle Physics and Astrophysics, CH-8093 Z\"{u}rich, Switzerland}
	\affiliation{Paul Scherrer Institut, CH-5232 Villigen PSI, Switzerland}	
	
	\author{D.~Ries}
	\affiliation{Department of Chemistry - TRIGA site, Johannes Gutenberg University Mainz, 55128 Mainz, Germany}
	
	\author{S.~Roccia}
	\affiliation{Universit\'e Grenoble Alpes, CNRS, Grenoble INP, LPSC-IN2P3, 38026 Grenoble, France}
	\affiliation{Institut Laue-Langevin, CS 20156 F-38042 Grenoble Cedex 9, France}
	
	\author{K.~U.~Ross}
	\affiliation{Department of Chemistry - TRIGA site, Johannes Gutenberg University Mainz, 55128 Mainz, Germany}
	
	\author{D.~Rozpedzik}
	\affiliation{Marian Smoluchowski Institute of Physics, Jagiellonian University, 30-348 Cracow, Poland}
	
	\author{P.~Schmidt-Wellenburg}
	\email[Corresponding author: ]{philipp.schmidt-wellenburg@psi.ch}
	\affiliation{Paul Scherrer Institut, CH-5232 Villigen PSI, Switzerland}
	
	\author{A.~Schnabel}
	\affiliation{Physikalisch-Technische Bundesanstalt, D-10587 Berlin, Germany}
	
	\author{N.~Severijns}
	\affiliation{Instituut voor Kern- en Stralingsfysica, University of Leuven, B-3001 Leuven, Belgium}
	
	\author{B.~Shen}
	\affiliation{Department of Chemistry - TRIGA site, Johannes Gutenberg University Mainz, 55128 Mainz, Germany}
	
	\author{R.~Tavakoli~Dinani}
	\affiliation{Instituut voor Kern- en Stralingsfysica, University of Leuven, B-3001 Leuven, Belgium}
	
	\author{J.~A.~Thorne}
	\affiliation{Laboratory for High Energy Physics and Albert Einstein Center for Fundamental Physics, University of Bern, CH-3012 Bern, Switzerland}
	
	\author{R.~Virot}
	\affiliation{Universit\'e Grenoble Alpes, CNRS, Grenoble INP, LPSC-IN2P3, 38026 Grenoble, France}
	
	\author{N.~Yazdandoost}
	\affiliation{Department of Chemistry - TRIGA site, Johannes Gutenberg University Mainz, 55128 Mainz, Germany}
	
	\author{J.~Zejma}
	\affiliation{Marian Smoluchowski Institute of Physics, Jagiellonian University, 30-348 Cracow, Poland}
	
	\author{G.~Zsigmond}	
	\affiliation{Paul Scherrer Institut, CH-5232 Villigen PSI, Switzerland}

	\begin{abstract}
		Magnetic Johnson-Nyquist noise (JNN) originating from metal electrodes, used to create a static electric field in neutron electric-dipole-moment (nEDM) experiments, may limit the sensitivity of measurements. We present here the first dedicated study on JNN applied to a large-scale long-measurement-time experiment with the implementation of a co-magnetometry. In this study, we derive surface- and volume-averaged root-mean-square normal noise amplitudes at a certain frequency bandwidth for a cylindrical geometry. In addition, we model the source of noise as a finite number of current dipoles and demonstrate a method to simulate temporal and three-dimensional spatial dependencies of JNN\@. The calculations are applied to estimate the impact of JNN on measurements with the new apparatus, n2EDM, at the Paul Scherrer Institute. We demonstrate that the performances of the optically pumped $^{133}$Cs magnetometers and $^{199}$Hg co-magnetometers, which will be used in the apparatus, are not limited by JNN\@. Further, we find that in measurements deploying a co-magnetometer system, the impact of JNN is negligible for nEDM searches down to a sensitivity of $4\,\times\,10^{-28}\,e\cdot{\rm cm}$ in a single measurement; therefore, the use of economically and mechanically favored solid aluminum electrodes is possible. \\
	\end{abstract}	
	\maketitle

	\section{Introduction} \label{Introduction}
	The search for a permanent electric dipole moment of the neutron (nEDM) has been an important topic in fundamental physics research since the 1950s\,\cite{Smith1957}. These experiments have been carried out by comparing the Larmor precession frequencies of neutron spins ($f_{\rm{n}}$) under static uniform parallel and anti-parallel electric and magnetic fields, using the Ramsey technique of separated oscillating fields\,\cite{Ramsey1950, Purcell1950}. For an accurate and precise measurement of $f_{\rm{n}}$, controlling the stability and uniformity as well as reducing the noise of the magnetic field in the apparatus are of paramount importance. A potential source of magnetic-field disturbance is Johnson-Nyquist noise (JNN)\,\cite{Johnson1928, Nyquist1928}, originating from the thermal motion of electrons in metal components within the experimental apparatus. 
	
	Johnson-Nyquist noise was originally observed as a ``random variation of potential between the ends of a conductor''\,\cite{Johnson1928}. The same underlying effect, random thermal motion of charge carriers, also results in fluctuation of the electromagnetic field near a conductor. Magnetic JNN is relevant in various research domains, all related to measurements with highest precision. The first published numerical analysis of JNN came from the bio-magnetic measurements\,\cite{Varpula1984} using superconducting quantum interference devices (SQUIDs). Johnson-Nyquist noise often exceeds the intrinsic noise of modern high-sensitivity detectors such as SQUIDs\,\cite{Varpula1984, Clem1987, Nenonen1996, Koerber2018} and high-density alkali atomic magnetometers\,\cite{Allred2002}. More recently, it was observed that these magnetic-near-field fluctuations induce spin-flip processes, and in turn are a crucial element of decoherence in magnetic traps which limits the trapping lifetime of atoms\,\cite{Jones2003, Harber2003, Lin2004, Rekdal2004, Henkel2005, Emmert2009}. In addition, relaxations of spin states in the presence of magnetic fluctuations were also studied in the context of magnetic resonance force microscopy and quantum computation\,\cite{Sidles2003}. In EDM measurements, following the requirement of sensitivity enhancement, potential constraints from JNN received extensive attentions in the last few decades\,\cite{Lamoreaux1999, Munger2005, Amini2007, Rabey2016}. Johnson-Nyquist noise from metal parts in high precision experiments impose a limit on the measurement sensitivity. Quantifying the impact of JNN in the design of the n2EDM experiment\,\cite{Abel2019a, Ayres2021} to search for an nEDM at the Paul Scherrer Institute (PSI) in Villigen, Switzerland, with the PSI ultracold-neutron (UCN) source\,\cite{Lauss2014, Bison2020} motivates the presented research. 
	
	As in the past, the challenges in measuring nEDM, $d_{\rm n}$, are the increase of the statistical sensitivity and the corresponding control of systematic effects. New sources of UCN worldwide\,\cite{Bison2017, Ito2018} improve the statistical sensitivity, $\sigma_{d_{\rm n}} \propto N^{-1/2}$, by increasing the number of neutrons available after storage, $N$. Nevertheless, owing to systematic effects, e.g.\ random frequency shifts due to possible magnetic-field noise or drifts, the improvement in pure counting statistics might be compromised. This study investigates the impact of JNN on nEDM experiments, focusing especially on the n2EDM spectrometer\,\cite{Abel2019a, Ayres2021} currently under construction at the PSI.
	
\begin{figure}[t!]
		\centering
		\includegraphics[width=.95\linewidth]{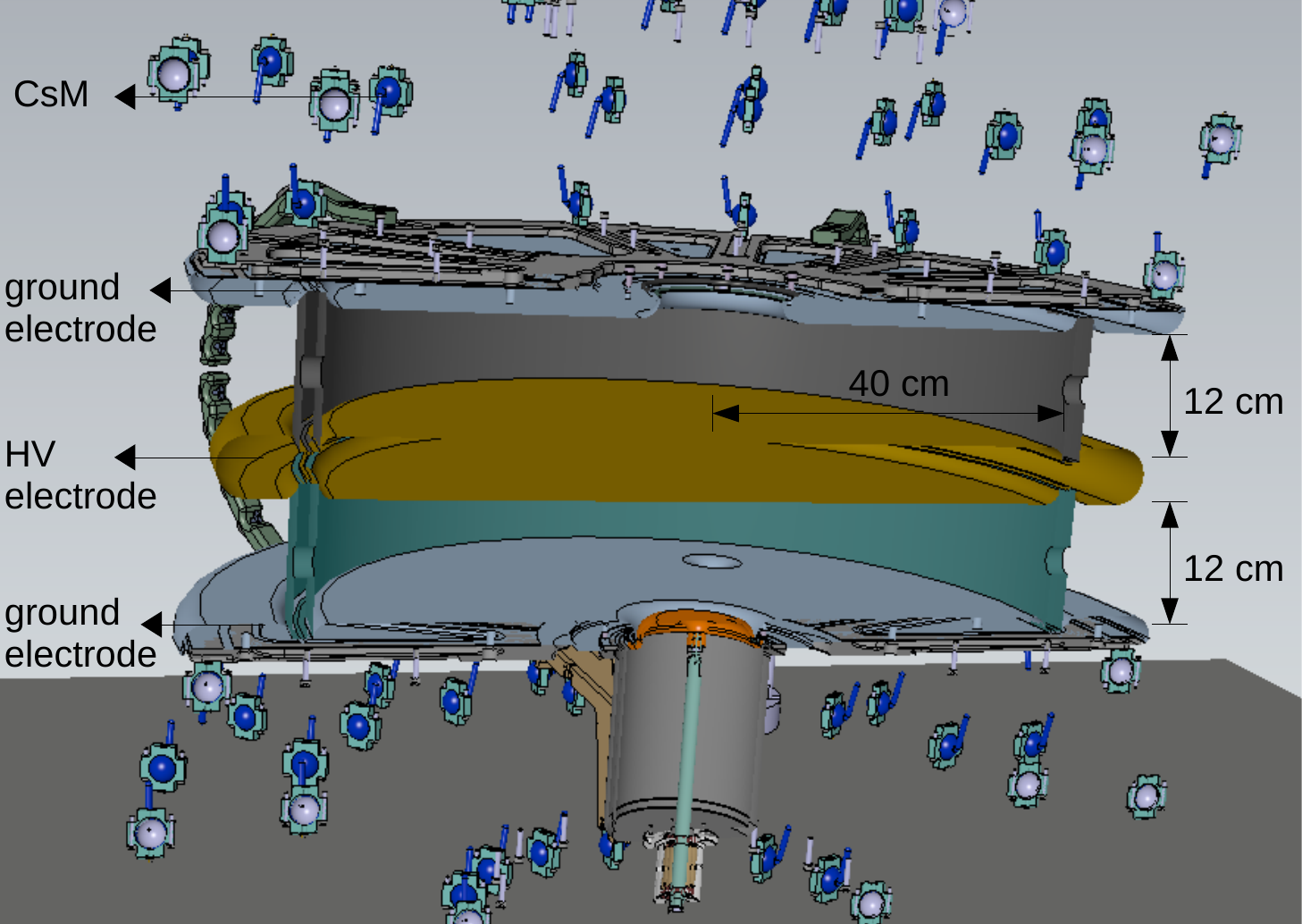}
		\caption{Cross-sectional drawing of the central storage chambers of the n2EDM apparatus located inside the vacuum tank. Included are only components relevant to this study. For a better visibility, the support structure of the Cs magnetometers (CsM) and the top UCN shutter are omitted. The electrodes are separated by polystyrene insulating rings.}
		\label{Fig_n2EDM}
	\end{figure}

	The n2EDM apparatus features two cylindrical storage chambers, $12$\,cm in height and $40$\,cm in radius, stacked vertically. The two chambers share the central plane, an electrode to which a high voltage of up to 200~kV can be applied. The top and bottom of the cylinder pair are closed with grounded plates. Figure~\ref{Fig_n2EDM} shows a simplified drawing of the central storage chambers with the components relevant to this paper, whereas detailed descriptions and schematics of the experimental apparatus can be found in Figs.\,1 and 2 of Ref.\,\cite{Abel2019a}. The three electrodes are made of aluminum plates, whose surfaces pointing towards the inside of the UCN storage chambers are coated with diamond-like carbon (DLC)\,\cite{Atchison2005, Atchison2006} in order to optimize the UCN reflection properties. 
	
	Thermal motion of charge carriers in the bulk aluminum results in magnetic JNN, which might affect the sensitivity of the magnetometers in its vicinity. In the following, we investigate the effects of JNN on the cesium magnetometers~(CsM)\,\cite{Weis2005, Groeger2006}, glass bulbs filled with saturated $^{133}$Cs vapor positioned around the precession chambers, the UCN and the mercury co-magnetometers~(HgM), polarized $^{199}$Hg atoms occupying the same volumes as the UCN in the chambers and read out by resonant light beams\,\cite{Green1998, Baker2014, Ban2018}. 
	
	\section{Spectrum of Magnetic Johnson-Nyquist Noise}
	The relevant magnetic JNN spectrum was first analytically derived for research in biophysical applications\,\cite{Varpula1984, Nenonen1996}. There it is shown that for an infinite-slab conductor of thickness $a$ and conductivity $\sigma$ at a temperature $T$, the normal component of the amplitude spectral density~(ASD) found in a distance $d$ to the conductor surface within a finite frequency interval $\Delta f$, with the $z$-axis defined to be perpendicular to the conductor surface, can be written as\,\cite{Varpula1984}
	\begin{equation} 
		\begin{aligned}
			\mathcal{B}_{z}(d, f) &= \sqrt{\overline{B_{z}(d, f)^{2}}/\Delta f} \\
			&= \mu_{0} \sqrt{\frac{2 \sigma k_{{\rm B}} T}{\pi}} \left[ \int_{0}^{\infty} R(\rho,a,\sigma,f) e^{-2\rho d} \rho \diff{\rho} \right]^{1/2}, \\
		\end{aligned}	
		\label{Eq_RmsLsdZ}
	\end{equation}
	where $k_{{\rm B}}$ is the Boltzmann constant, $\rho$ is the radial component of the infinite conductor, and $B_{z}(d, f)$ is the magnetic-field amplitude normal to the surface at a given frequency $f$\,\footnote{Compare to Eq.\,(38) of Ref.\,\cite{Varpula1984}, where variables $z$, distance, and $t$, thickness, were replaced with $d$ and $a$ in our paper to avoid confusion with other variables used in the remainder of the article.}. $R(\rho,a,\sigma,f)$ in Eq.\,\eqref{Eq_RmsLsdZ} is a function of conductor properties, $a$ and $\sigma$, at a given frequency $f$, with the original expression defined in Eq.\,(39) of Ref.\,\cite{Varpula1984}. For the horizontal components, due to symmetry and Maxwell's equations, one can infer that\,\cite{Varpula1984}
	\begin{equation} 
		\mathcal{B}_{x}(d, f) = \mathcal{B}_{y}(d, f) = \dfrac{1}{\sqrt{2}} \mathcal{B}_{z}(d, f).
		\label{Eq_RmsLsdX}
	\end{equation}
	
\begin{figure}[h!]
		\centering
		\includegraphics[width=.95\linewidth]{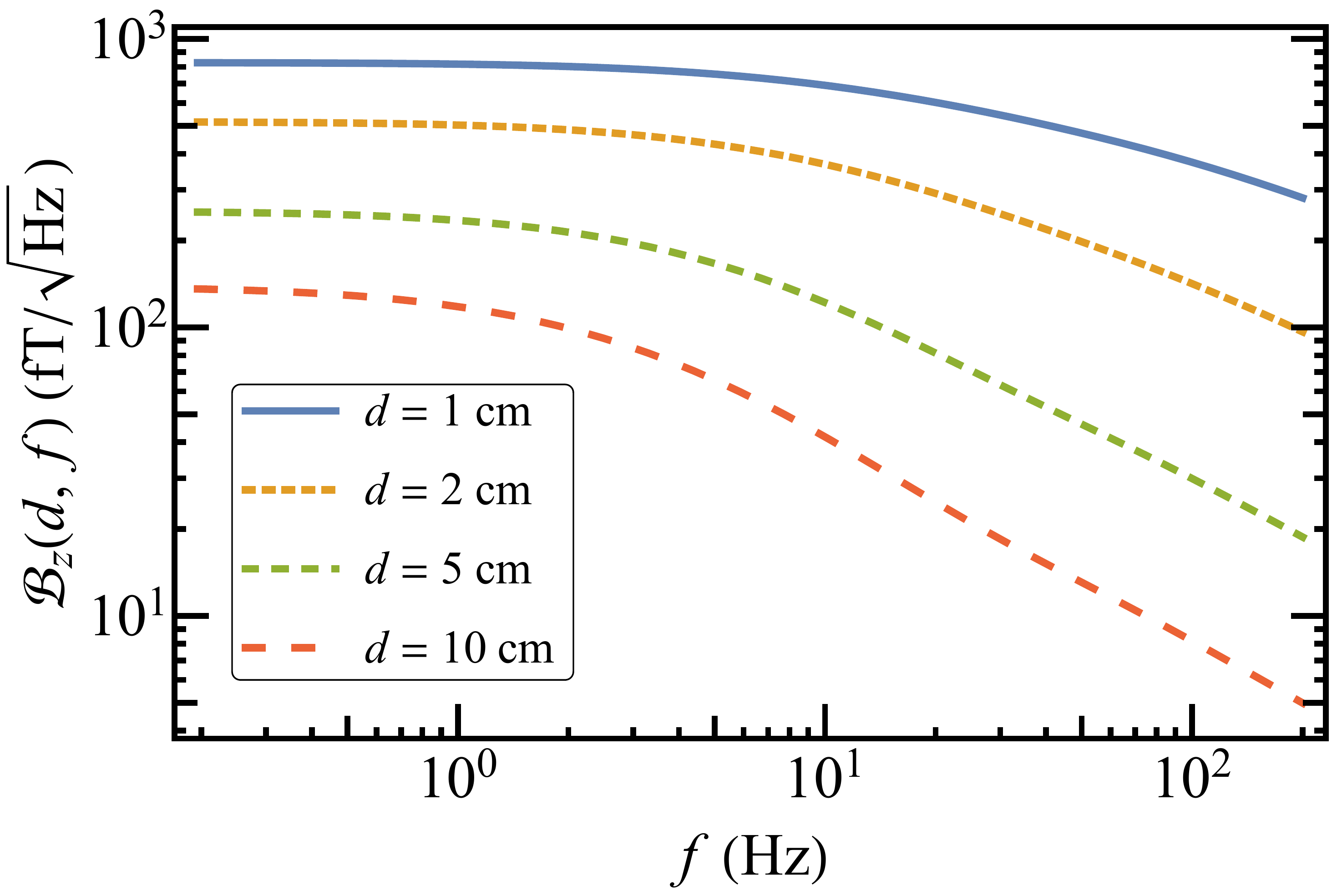}
		\caption{Normal amplitude spectral density of a 2.5-cm-thick aluminum ($\sigma=3.77 \times 10^7$~S/m) infinite-slab conductor at various distances under 20\,\si{\degreeCelsius}.}
		\label{Fig_Bz}
	\end{figure}

	All three components of the noise spectrum depend on the normal distance between the point of interest and the surface of the infinite slab. Figure~\ref{Fig_Bz} shows the normal ASD of a 2.5-cm-thick aluminum infinite-slab conductor, $\sigma=3.77 \times 10^7$~S/m, under 20\,\si{\degreeCelsius} at various distances, using Eq.\,\eqref{Eq_RmsLsdZ}. The spectral line flattens at low frequencies towards a constant value equal to the root-mean-square (RMS) limit for $f\rightarrow0$. With increasing frequency, the noise amplitude decreases due to self-damping of high-frequency noise within the conductor slab.
	
	To verify these noise spectra, measurements of the magnetic-field noise created by an aluminum sheet were carried out in the magnetically shielded room BMSR-2 at the Physikalisch-Technische Bundesanstalt (PTB), Berlin, Germany\,\cite{Bork2001}, using the 304-channel SQUID-vector-magnetometer system\,\cite{Thiel2005}. The system is based on nineteen modules placed on a hexagonal grid with each comprising of sixteen SQUIDs placed at various vertical planes. A total of 304 SQUIDs each with a 7~mm effective pickup-coil diameter permit to calculate all three vector components of the magnetic field on three measurement planes. The noise of a sheet made of 99.5\% aluminum with dimensions of 1.3~m\,$\times$\,1~m $\times$ 0.5~mm was measured. In the first measurement, the sheet was placed to touch the flat bottom of the dewar cryostat, resulting in a minimal distance of 27.5~mm between sample and SQUIDs, due to the cold-warm distance of the dewar. Another measurement was carried out by adding a wood plate in between aluminum sheet and dewar to provide an additional 7.5~mm distance to the dewar bottom.
	In each measurement, seven SQUIDs at different heights from two central modules were used, with four measuring the vertical field component w.r.t.\ the probe and three measuring the horizontal field component. 
	
\begin{figure}
		\centering
		\begin{subfigure}[h!]{.45\textwidth}
			\includegraphics[width=\linewidth]{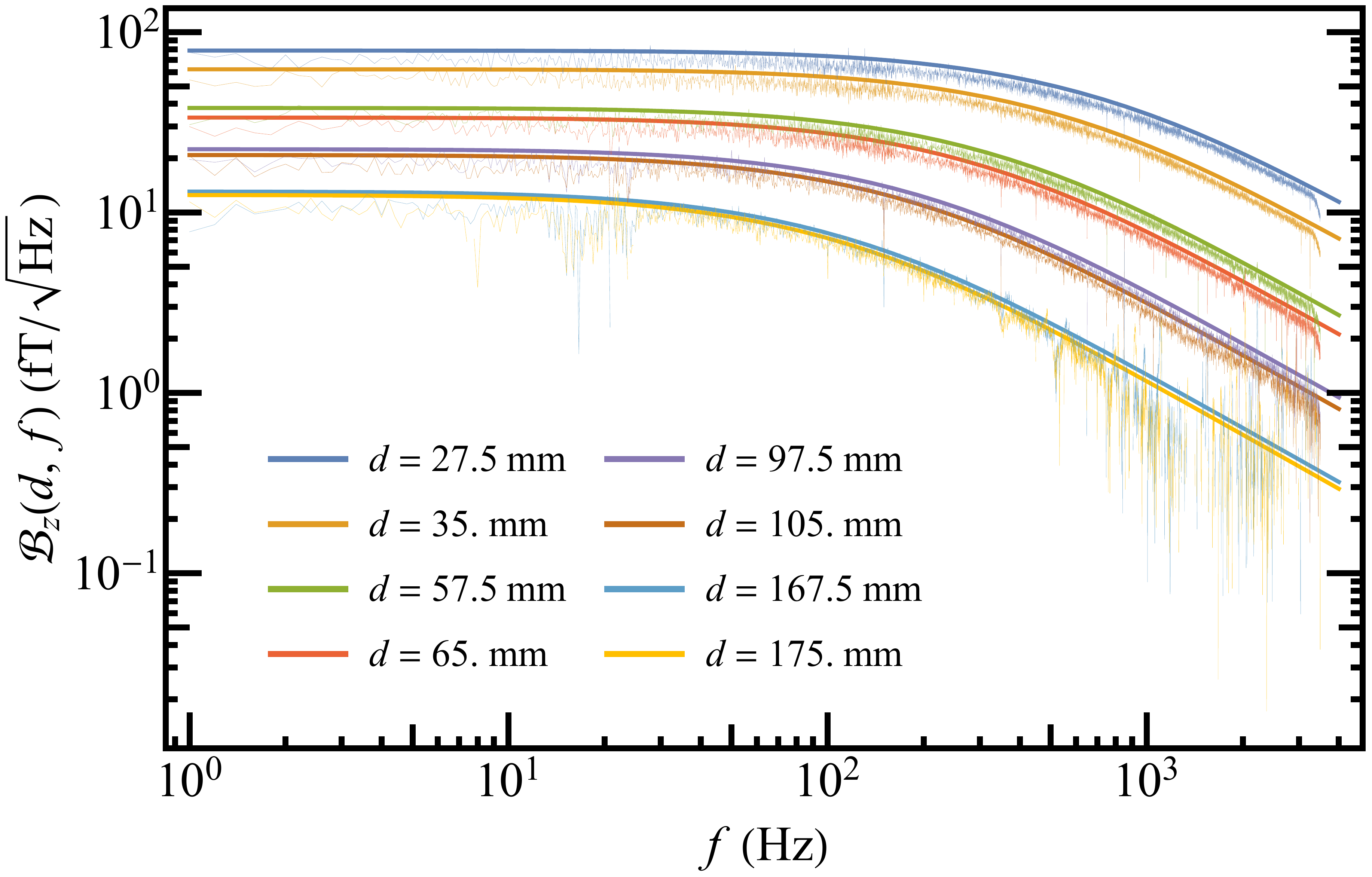}
			\caption{} \label{SubFig_BzPtb}
		\end{subfigure}
		\begin{subfigure}[h!]{.45\textwidth}
			\includegraphics[width=\linewidth]{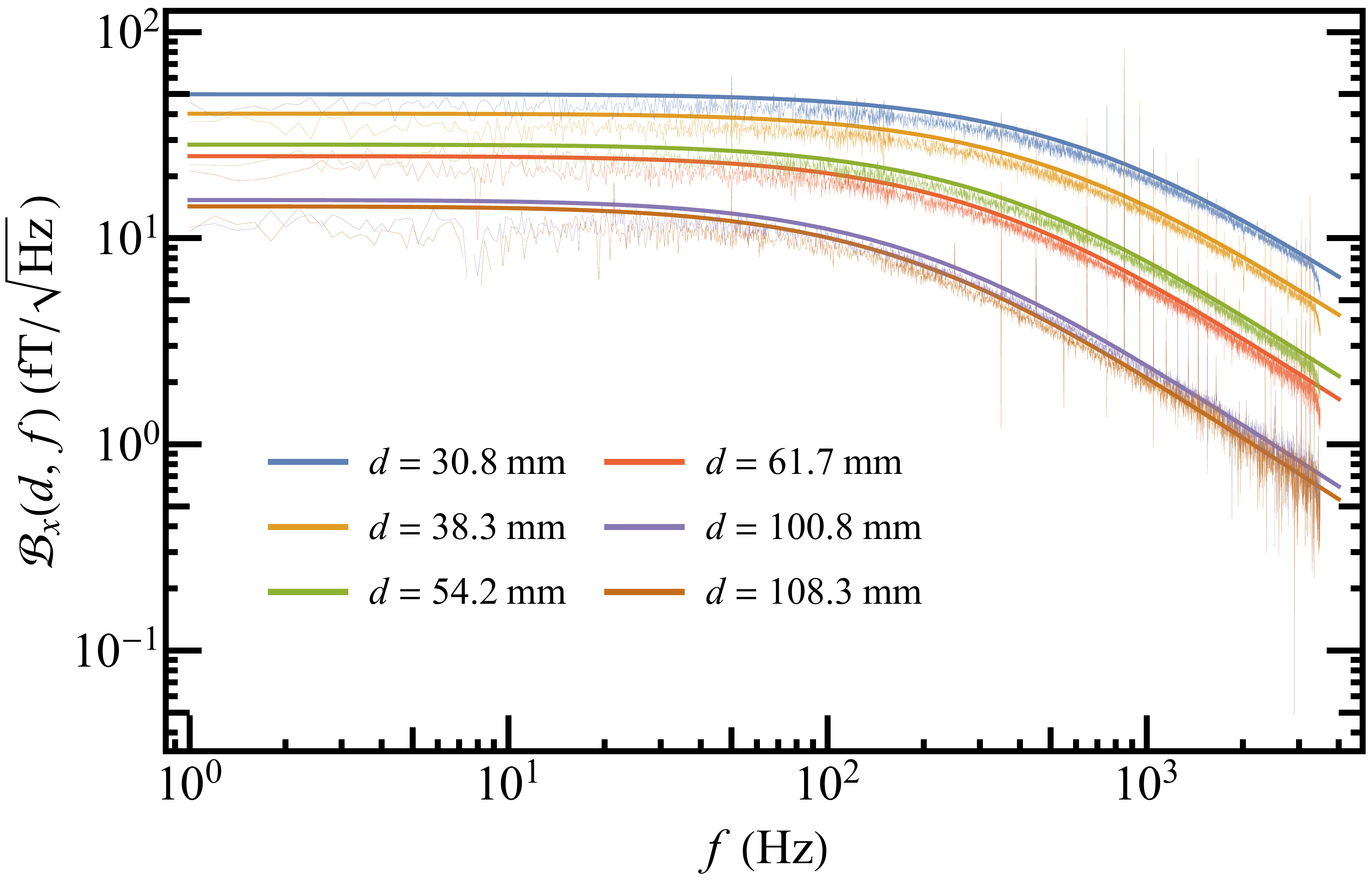}
			\caption{} \label{SubFig_BxPtb}
		\end{subfigure}
		\caption{(a) Vertical and (b) horizontal magnetic-field noise component of a 0.5-mm-thick aluminum sheet measured by the SQUID system at PTB w.r.t.\ the theoretical ASD at various distances. The spectra were averaged over 5~s samples from 300~s measurements. The ASD decreases with increasing distance.}
		\label{Fig_BPtb}
	\end{figure}

	Figures~\ref{SubFig_BzPtb} and \ref{SubFig_BxPtb} are the combined results from the two measurements, with or without the wood plate, with the former showing the vertical field component and the latter displaying the horizontal component. 
	The spectra are results from 300~s measurements averaged over 5~s samples. The background noise measured without aluminum sheet was subtracted. The increase in noise below $\SI{2}{fT/\sqrt{Hz}}$ is due to the SQUID white noise. Mechanical vibrations influence the measurement between 5-25~Hz at the level of $\SI{10}{fT/\sqrt{Hz}}$. Comparing between the measured spectra and the theoretical ASD calculated with Eqs.\,\eqref{Eq_RmsLsdZ} and \eqref{Eq_RmsLsdX}, with $\sigma=3.77 \times 10^7$~S/m and $T=22$~\si{\degreeCelsius}, discrepancies up to $\sim 15\%$ are found. The uncertainties on $\sigma$ and $T$ are not enough to explain the differences. Nenonen, Montonen and Katila\,\cite{Nenonen1996} pointed out that correlation of JNN within pickup coils with finite surface areas should be taken into account if the pickup-coil diameter is larger than the measurement distance (see Sec.\,\ref{Sec_AnalyticalWithMC} for details). In our measurements, the ratios between the pickup-coil diameter and the distances were all smaller than one third; thus, the correlation is considered negligible (see Fig.\,\ref{Fig_DiskIntegral}). Additionally, considering the dimensions of the aluminum sheet, the pick-up coils and the measurement separations, we can approximate the aluminum sheet as an infinite conductor. 
	
	Comparisons between theoretical calculations and noise measurements on copper\,\cite{Varpula1984} and aluminum\,\cite{Nenonen1996} conductors using first-order gradiometers have been performed in bio-magnetic researches. It is reported in Ref.\,\cite{Varpula1984} that the agreement lies within uncertainties of $\pm 15\%$. Our measurements performed with a magnetometer confirm the theoretical spectrum in a non-differential manner to a fairly good level, and verify the relation between transverse and normal noise components (Eq.\,\eqref{Eq_RmsLsdX}). 
	The reason for the small disagreement is still unclear; nonetheless, by using the theoretical ASD, we guarantee that the noise will not be underestimated. 	
	
	For JNN studies in EDM experiments, a principle simplification can be made assuming that the frequency of the fluctuating field is low enough such that the eddy currents generated in the bulk material can be neglected. This is stated as the \textit{static approximation} by Lamoreaux\,\cite{Lamoreaux1999}, which is valid when the thickness of the conductor is smaller than the skin depth
	\begin{equation}
		\lambda=\sqrt{\frac{1}{\pi\mu\sigma f}},
	\end{equation}  
	where $f$ is the fluctuation frequency. In the context of EDM experiments, this corresponds to approximately the reciprocal of the spin-coherence time, $T_2$. In the n2EDM experiment, the free-spin-precession period for a single measurement will be $\Delta t \sim$ 200~s, approximately two times the spin-coherence time of mercury and a fraction of the spin-coherence time of neutrons. Hence, we assume $f^{-1} =\Delta t \sim \SI{200}{s} \approx T_{2,{\rm Hg}}$ will be the free-spin-precession period for a single measurement. At 5~mHz, $\lambda \sim$ 116~cm, so the low-frequency assumption can be applied safely for conductors with a thickness of less than 10~cm. 
	
	\section{Magnetic-field fluctuation observed by field-sensing particles}
	During an nEDM-measurement cycle, $^{199}$Hg atoms occupying the same volumes as UCN, are used as co-habiting magnetometers (HgM)\,\cite{Green1998, Baker2014, Ban2018}. As the HgM and the UCN measure the magnetic field simultaneously, the ratio of the two precession frequencies, $f_{{\rm n}}/f_{{\rm Hg}}$, is robust against magnetic-field changes. Nonetheless, the two spin species sample the magnetic field differently. The UCN sample the field adiabatically and have a negative center-of-mass offset whereas the $^{199}$Hg atoms sample the field non-adiabatically\,\cite{Abel2019}. For a nominal field of $B_0=1$\si{\micro T}, we investigate the degree to which the effects of JNN can be controlled when taking the frequency ratio of two ensembles within one precession chamber.

	\subsection{Analytical derivation of spatial properties} \label{Sec_AnalyticalDerivation}
	In the first step to calculate the RMS magnetic-field noise sensed by the particles, it is useful to derive the spatial correlation of JNN at different locations within the volume. For this purpose we calculated the magnetic noise originating from thermal noise currents by dividing a volume conductor into infinitesimal cuboidal elements, $\Delta V = \Delta x \Delta y \Delta z$, similar to the seminal calculation in Refs.\,\cite{Varpula1984, Nenonen1996}. There an {\it equivalent current dipole} for the volume element is defined, whose component $P_{\alpha} = I_{\alpha} \Delta \alpha$  $(\alpha = x, y, z)$ in the direction $\alpha$ is the product of this short-circuit current and the finite size of the element. Following this concept, the source of thermal magnetic noise is represented by a great number of randomly oriented current dipoles on the surface of the conductor. 
	
	We consider an infinite conductor and assume its surface is an $x-y$ plane on the reference of the vertical coordinate $z=0$. A current dipole element on an infinitesimal surface area $\diff{s}$ located at $(\boldsymbol{x},0)$, where $\boldsymbol{x}$ now denotes a two-dimensional vector on the $x-y$ plane, is written as $\boldsymbol{I}(\boldsymbol{x})\diff{s}$, where the $z=0$ component is omitted for simplicity. The magnetic field created by this dipole at a point $(\boldsymbol{r}, z)$ can be calculated, according to the Biot-Savart law, as 
	\begin{equation}
		\diff{\boldsymbol{B}} = \frac{\mu_0}{4\pi} \frac{\boldsymbol{I}(\boldsymbol{x})\diff{s} \times \boldsymbol{k}}{d^{2}},
	\end{equation}
	where 
	\begin{equation}
		\boldsymbol{k} = \frac{\boldsymbol{r}-\boldsymbol{x}}{d}  + \frac{z}{d} \, \hat{e_z}
	\end{equation}
	is the unit directional vector pointing from $(\boldsymbol{x}, 0)$ to $(\boldsymbol{r}, z)$, and
	\begin{equation}
		d = \sqrt{{\left| \boldsymbol{r}-\boldsymbol{x} \right|}^2 + z^2}
	\end{equation}
	is the distance between the dipole and the observation point. Now, we obtain the normal component $\hat{e}_z$ of the field 
	\begin{align}
		\diff{B_z}(\boldsymbol{r}, z) &= \diff{\boldsymbol{B}} \cdot \hat{e}_z \nonumber \\
		&= \frac{\mu_0}{4\pi} \left( \boldsymbol{I}(\boldsymbol{x}) \diff{s} \times \frac{\boldsymbol{r}-\boldsymbol{x}}{d^3} \right) \cdot \hat{e}_z \nonumber \\
		&= \frac{\mu_0}{4\pi} \left( \hat{e}_z \times \boldsymbol{I}(\boldsymbol{x}) \diff{s} \right) \cdot \frac{\boldsymbol{r}-\boldsymbol{x}}{d^3} \nonumber \\
		&= \frac{\mu_0}{4\pi} \boldsymbol{F}(\boldsymbol{r}-\boldsymbol{x}, z) \cdot \boldsymbol{\mathcal{I}}(\boldsymbol{x}) \diff{s},  
	\end{align}
	with 
	\begin{equation}
		\boldsymbol{F}(\boldsymbol{x}, z) \equiv \frac{\boldsymbol{x}}{\left( \left| \boldsymbol{x} \right|^2 + z^2 \right)^{3/2}}
	\end{equation}
	and 
	\begin{equation}
		\boldsymbol{\mathcal{I}}(\boldsymbol{x}) \equiv \hat{e}_z \times \boldsymbol{I}(\boldsymbol{x}) 
	\end{equation}
	being the rotated current component transformed from the triple product.

	\subsubsection{Variance of a disk-averaged field}
	Consider a disk parallel to the conductor, which has a radius $R$ and is located at a normal distance $z$ above the conductor. The average normal magnetic field over this disk generated by thermal noise in a finite element $\diff{s}$ from the conductor can be written as 
	\begin{align}
		\diff{\bar{B}_z}(R, z) &\equiv  \int_{S_R}   \frac{\diff^2 r}{\pi R^2} \, \diff{B_z}(\boldsymbol{r}, z) \nonumber \\
		&=\frac{\mu_0}{4\pi} \int_{S_R} \frac{\diff^2 r}{\pi R^2} \,  \boldsymbol{F}(\boldsymbol{r}-\boldsymbol{x}, z) \cdot \boldsymbol{\mathcal{I}}(\boldsymbol{x}) \diff{s} \nonumber \\
		&= \frac{\mu_0}{4\pi} \, \boldsymbol{\mathcal{I}}(\boldsymbol{x}) \diff{s} \cdot \bar{\boldsymbol{M}}(\boldsymbol{x}, R, z),
	\end{align}
	where  
	\begin{equation}
		\bar{\boldsymbol{M}}(\boldsymbol{x}, R, z) \equiv  \int_{S_R} \frac{\diff^2 r}{\pi R^2} \boldsymbol{F}(\boldsymbol{r}-\boldsymbol{x}, z)
	\end{equation}
	is the average over the disk.
	For an infinite conductor, we integrate over all dipoles 
	\begin{equation}
		\bar{B}_z(R, z) = \int \diff{\bar{B}_z}(R, z). 
	\end{equation} 	
	The variance of this surface average is then calculated as 
	\begin{equation}
		\begin{split}
			&\left\langle\bar{B}_z(R, z)^2\right\rangle =
			\left( \frac{\mu_0}{4\pi} \right)^2 \int \diff{s} \int \diff{s'} \\
			&\qquad \left\langle \left( \boldsymbol{\mathcal{I}}(\boldsymbol{x}) \cdot  \bar{\boldsymbol{M}}(\boldsymbol{x}, R, z) \right) 
			\left( \boldsymbol{\mathcal{I}}(\boldsymbol{x'}) \cdot \bar{\boldsymbol{M}}(\boldsymbol{x'}, R, z) \right) \right\rangle.
		\end{split}
	\end{equation}
	
	As shown in Eq.\,(1) in Ref.\,\cite{Varpula1984}, based on Nyquist's theorem, 
	\begin{equation}
		\left\langle \boldsymbol{\mathcal{I}}(\boldsymbol{x}) \boldsymbol{\mathcal{I}}(\boldsymbol{x'}) \right\rangle 
		= 4 \sigma k_{{\rm B}} T \Delta f a \delta(\boldsymbol{x}-\boldsymbol{x'}),   
	\end{equation}
	where the conductor properties are identical to those indicated in Eq.\,\eqref{Eq_RmsLsdZ}. With the change of variables and further derivations, the variance of the surface-averaged field can be expressed as 
	\begin{align}
		\left\langle \bar{B}_z(R, z)^2 \right\rangle &= \frac{\mathcal{C}\pi}{2 z^2} \mathfrak{I}\left(\frac{R}{z}\right) \nonumber \\ 
		&= \left\langle B_z(\boldsymbol{r}, z)^2 \right\rangle \mathfrak{I} \left(\frac{R}{z}\right) \nonumber \\
		&= \left\langle B_z(0, z)^2 \right\rangle \mathfrak{I} \left(\frac{R}{z}\right), 
		\label{Eq_SurAvgNoise}
	\end{align} 
	normalized to the variance of the single-point field at a random location of distance $z$, $\left\langle B(\boldsymbol{r}, z)^2 \right\rangle = \left\langle B(0, z)^2 \right\rangle$, with
	\begin{equation}
		\mathcal{C} \equiv \left( \frac{\mu_0}{4\pi} \right)^2 4 \sigma k_{{\rm B}} T \Delta f a.
	\end{equation}
	$\mathfrak{I}(R/z)$ is an integration over three two-dimensional vectors calculated as  
	\begin{align}
		\mathfrak{I}(\xi) &\equiv \frac{2}{\pi^3}\xi^4 \int_{S_1} \diff^2 u \int_{S_1} \diff^2 v \int \diff^2 X \nonumber \\ 
		&\qquad \frac{\boldsymbol{u}-\boldsymbol{X}}{\left( \xi^2 \left| \boldsymbol{u}-\boldsymbol{X} \right|^2 + 1 \right)^{3/2}} \cdot
		\frac{\boldsymbol{v}-\boldsymbol{X}}{\left( \xi^2 \left| \boldsymbol{v}-\boldsymbol{X} \right|^2 + 1 \right)^{3/2}}. \label{Eq_IntI}
	\end{align}
	$\boldsymbol{u}$ and $\boldsymbol{v}$ are two observation points on the disk, where an integration over a unit circle $S_1$ is performed, and $\boldsymbol{X}$ is a dipole on the conductor integrated from zero to infinity.

	\subsubsection{Variance of a cylinder-averaged field}
	In our case, we are interested in the average field observed within a cylinder of radius $R$ and height $H$ on the surface of the conductor,
	\begin{equation}
		\diff{\bar{\bar{B}}_z}(R, H) \equiv \int_0^H \frac{\diff{z}}{H} \int_{S_R} \frac{\diff^2 r}{\pi R^2} \, \diff{B_z}(\boldsymbol{r}, z).
	\end{equation}
	Physically, direct contact between a dipole and an observation point will result in a divergent magnetic field; hence, we regularize the integration by starting from a small distance $h$ ($h\ll H, R$) to the conductor surface 
	\begin{align}
		\diff{\bar{\bar{B}}_z}(R, H) &\approx \diff{\bar{\bar{B}}_z}(R, H, h) \nonumber \\
		&= \int_h^{h+H} \frac{\diff{z}}{H} \int_{S_R} \frac{\diff^2 r}{\pi R^2} \, \diff{B_z}(\boldsymbol{r}, z) \nonumber \\
		&= \frac{\mu_0}{4\pi} \int_h^{h+H} \frac{\diff{z}}{H} \int_{S_R} \frac{\diff^2 r}{\pi R^2} \boldsymbol{F}(\boldsymbol{r}-\boldsymbol{x}, z) \cdot \boldsymbol{\mathcal{I}}(\boldsymbol{x}) \diff{s}  \nonumber \\
		&= \frac{\mu_0}{4\pi} \boldsymbol{\mathcal{I}}(\boldsymbol{x}) \diff{s} \cdot \bar{\bar{\boldsymbol{M}}}(\boldsymbol{x}, R, H, h), 
	\end{align}
	with 
	\begin{equation}
		\bar{\bar{\boldsymbol{M}}}(\boldsymbol{x}, R, H, h) \equiv \int_{S_R} \frac{\diff^2 r}{\pi R^2} \bar{\boldsymbol{F}}(\boldsymbol{r}-\boldsymbol{x}, H, h)
	\end{equation}
	and 
	\begin{equation}
		\bar{\boldsymbol{F}}(\boldsymbol{x}, H, h) \equiv \int_h^{h+H} \frac{\diff{z}}{H} \boldsymbol{F}(\boldsymbol{x}, z).
	\end{equation}
	Similarly, the contributions of all dipoles are integrated over, $\bar{\bar{B}}_z(R, H, h) = \int \diff{\bar{\bar{B}}_z}(R, H, h)$, and the variance of the volume-averaged field can be carried out as 
	\begin{align}
		\left\langle \bar{\bar{B}}_z(R, H, h)^2 \right\rangle = \frac{\mathcal{C}}{\pi^2 R^2} \mathfrak{J} \left( \frac{H}{R}, \frac{h}{R}\right),
		\label{Eq_VolAvgNoise}
	\end{align}	
	where 
	\begin{align}
		\mathfrak{J} \left( \eta, \zeta \right) &\equiv \frac{1}{\eta^2} \int_{S_1} \diff^2 u \int_{S_1} \diff^2 v \int \diff^2 X \nonumber \\
		&\qquad \left\lbrace  \frac{\left(\eta + \zeta \right) \left(\boldsymbol{u}-\boldsymbol{X} \right)}{\left| \boldsymbol{u}-\boldsymbol{X} \right|^2 \left[ \left| \boldsymbol{u}-\boldsymbol{X} \right|^2 + \left(\eta + \zeta \right)^2 \right]^{1/2}} \right. - \nonumber \\ 
		&\qquad \qquad \left. \frac{\zeta \left( \boldsymbol{u}-\boldsymbol{X} \right)}{\left| \boldsymbol{u}-\boldsymbol{X} \right|^2 \left( \left| \boldsymbol{u}-\boldsymbol{X} \right|^2 + \zeta^2 \right)^{1/2}} \right\rbrace \cdot \nonumber \\ 
		&\qquad \left\lbrace  \frac{\left( \eta + \zeta \right) \left(\boldsymbol{v}-\boldsymbol{X} \right)}{\left| \boldsymbol{v}-\boldsymbol{X} \right|^2 \left[ \left| \boldsymbol{v}-\boldsymbol{X} \right|^2 + \left(\eta + \zeta \right)^2 \right]^{1/2}} \right. - \nonumber \\ 
		&\qquad \qquad \left. \frac{\zeta \left( \boldsymbol{v}-\boldsymbol{X} \right)}{\left| \boldsymbol{v}-\boldsymbol{X} \right|^2 \left(\left| \boldsymbol{v}-\boldsymbol{X} \right|^2 + \zeta^2 \right)^{1/2}} \right\rbrace. \label{Eq_IntJ}	
	\end{align}
	At the limit of $H\rightarrow 0$, the variance of the volume average reduces to the variance of the disk average at a distance $h$, which gives
	\begin{equation}
		\left\langle \bar{\bar{B}}_z(R, H\rightarrow0, h)^2 \right\rangle \approx \left\langle \bar{B}_z(R, h)^2 \right\rangle. 
		\label{Eq_VolLim}
	\end{equation}

	\subsection{Analytical derivation computed with Monte Carlo integration} \label{Sec_AnalyticalWithMC}
	The variances of surface and volume averages are important for practical purposes. To calculate the corresponding results, integrals in Eqs.\,\eqref{Eq_IntI} and \eqref{Eq_IntJ} were computed using the method of Monte Carlo integration\,\cite{Caflisch1998}. 
	
\begin{figure}[h!]
		\centering
		\includegraphics[width=.8\linewidth]{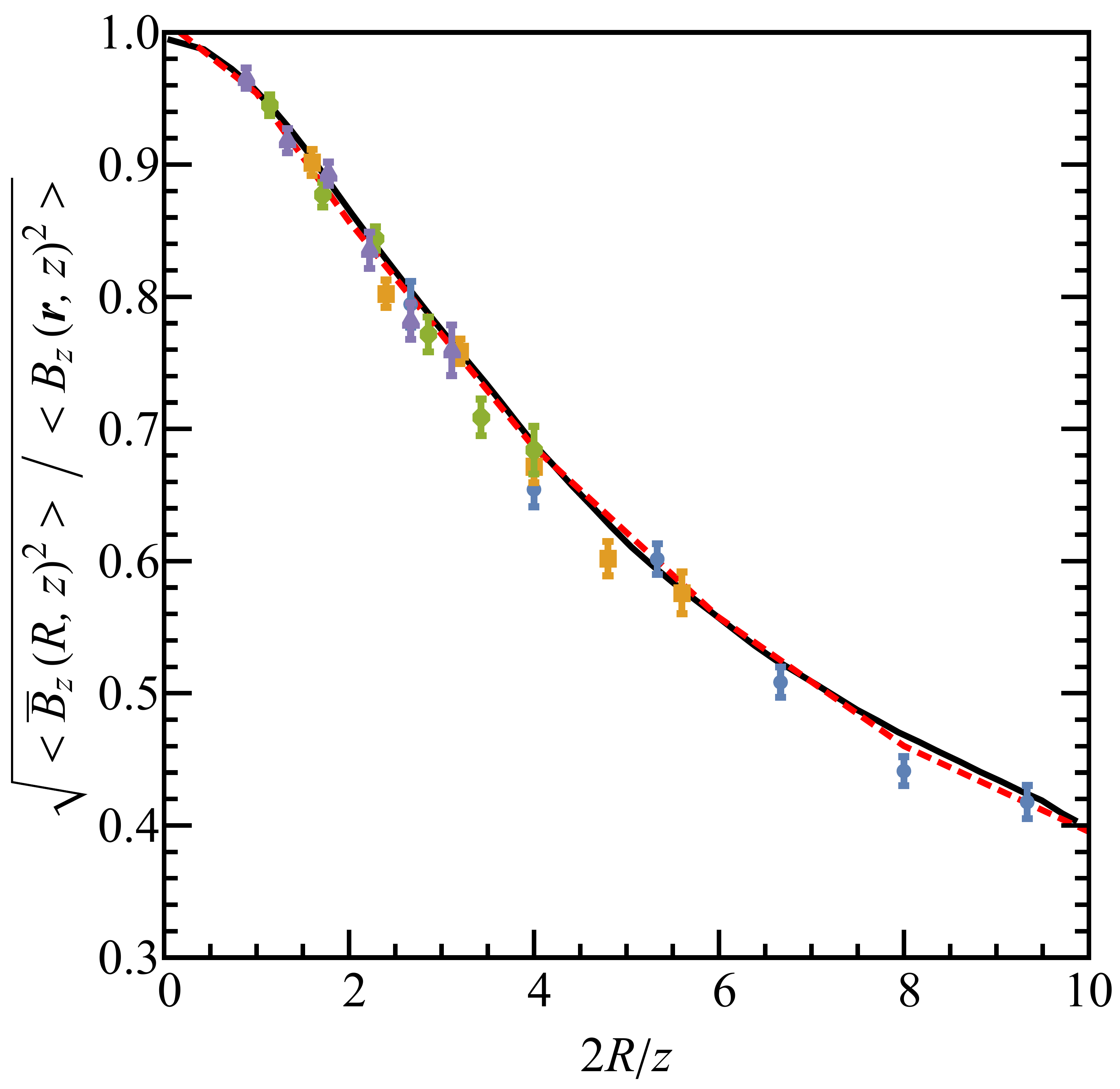}
		\caption{Root-mean-square normal noise averaged over a finite area w.r.t.\ a random single point $(\boldsymbol{r}, z)$. Comparison among results obtained with Monte Carlo integration based on analytical derivation (red dashed line), calculation shown in Ref.\,\cite{Nenonen1996} (black solid line) and numerical finite-element method computed at various distances (colored points).}
		\label{Fig_DiskIntegral}
	\end{figure}

	As described in Ref.\,\cite{Nenonen1996}, SQUID detectors used to measure magnetic fields have pickup coils with finite surface areas within which the correlation of JNN should be taken into account. Nenonen {\it et al.}\,\cite{Nenonen1996} calculated the magnetic noise observed by a single circular coil of diameter $d$ parallel to a conducting slab at a distance $z$, $B_{n,z}^{\rm coil}$, and plotted the ratio to the single-point spectral density, $B_{n,z}^{\rm coil}/B_{n,z}$, as a function of $d/z$, shown in Fig.\,4 of Ref.\,\cite{Nenonen1996}. $B_{n,z}^{\rm coil}$ and $B_{n,z}$ are the notations used in the original paper of Nenonen {\it et al.}, where $n$ in the subscript stands for JNN\@. They are equivalent to $\bar{B}_z(R, z)$ and $B_z(\boldsymbol{r}, z)$ in our study, respectively. The ratio $B_{n,z}^{\rm coil}/B_{n,z}$ is the same as $\mathfrak{I}\left(R/z\right)^{1/2}$ in Eq.\,\eqref{Eq_SurAvgNoise}. 
	We computed this integral $\mathfrak{I}\left(R/z\right)$ and compared it to the calculation made by Nenonen {\it et al.\ }shown in Fig.~\ref{Fig_DiskIntegral}. 
	The black solid line in the graph is the result from Ref.\,\cite{Nenonen1996}. 
	The red dashed line is our result using the Monte Carlo integration, averaged over thirty random numerical solutions. The other colored points were calculated with a numerical finite-element method which will be explained in Secs.\,\ref{Sec_FEM} and \ref{Sec_CfAnalyticalFEM}. 
	All methods agree with one another, and the remaining small deviations are inconsequential for our pragmatic intent. 
	
\begin{figure}[h!]
		\centering
		\includegraphics[width=.95\linewidth]{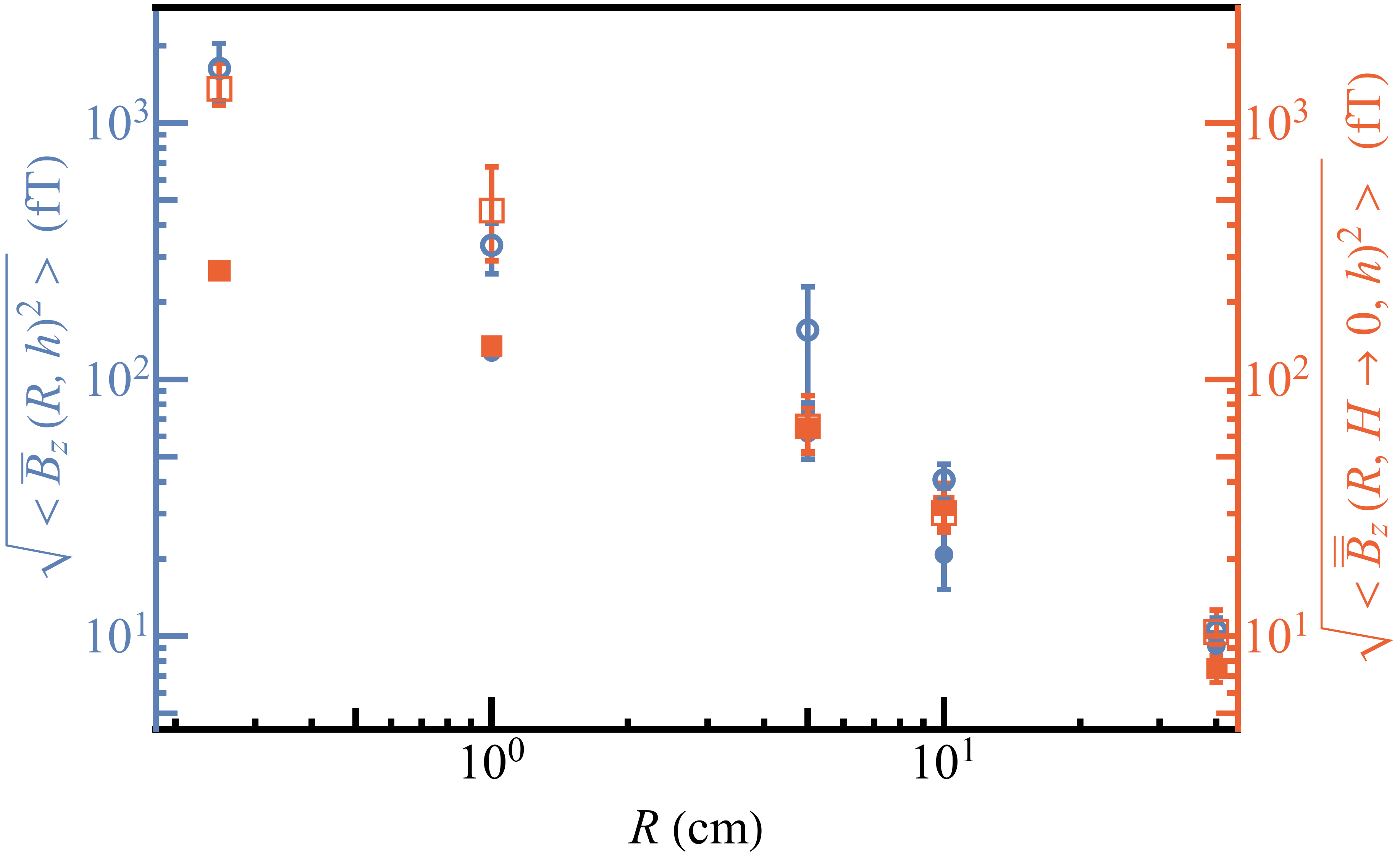}
		\caption{Comparison of RMS normal noise amplitude at a frequency bandwidth $\Delta f = 1/(2\!\times\!200\,{\rm s})$ between surface average (blue, circular) and volume average with an infinitesimal cylinder height (red, square). Two regularization distances, $h=$ \SI{2.5}{mm} or \SI{10}{\micro m}, are shown as filled or open data points, respectively.}
		\label{Fig_DiskVsCylinder}
	\end{figure}

	As described in Eq.\,\eqref{Eq_VolLim}, with reduction of cylinder height, the volume variance converges to the surface variance. Figure\,\ref{Fig_DiskVsCylinder} displays both $\left\langle \bar{\bar{B}}_z(R, H\rightarrow0, h)^2 \right\rangle^{1/2}$ (red, square) and $\left\langle \bar{B}_z(R, h)^2 \right\rangle^{1/2}$ (blue, circular) for various chamber radii. Integrations were performed using ten to fifty random solutions in the Monte Carlo method for various radii, where the mean values are plotted with their standard errors shown as error bars. 
	Filled data points were computed with a regularization distance $h=\SI{2.5}{mm}$, whereas open points were calculated with $h=\SI{10}{\micro\meter}$. 
	Both methods agree with each other which confirms the validity of the convergence of the volume calculation to the surface solution in the limit of zero chamber height ($H\rightarrow0$).   
	
\begin{figure}[h!]
		\centering
		\includegraphics[width=.95\linewidth]{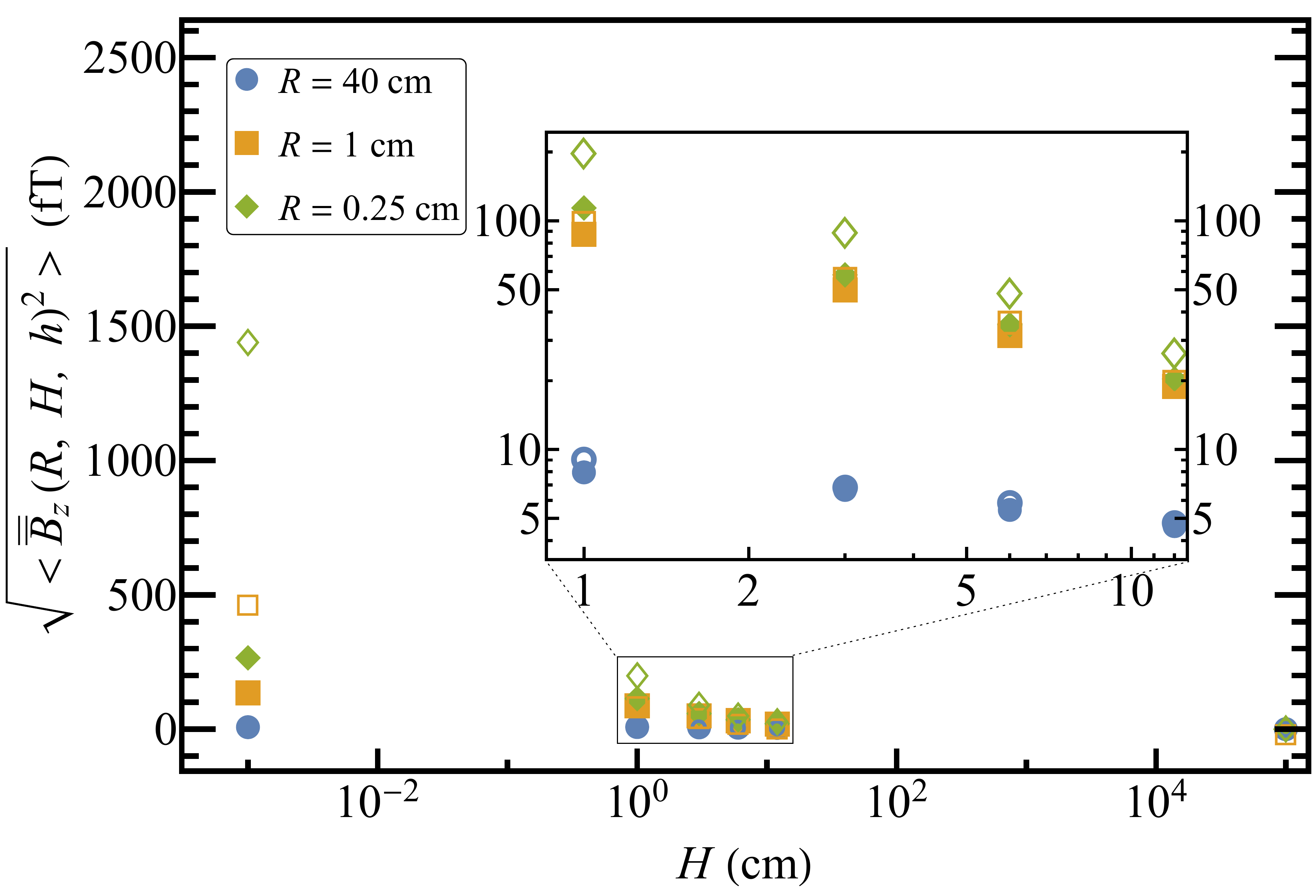}
		\caption{Volume-averaged RMS normal noise amplitude at a frequency bandwidth $\Delta f = 1/(2\!\times\!200\,{\rm s})$ with various cylinder dimensions. Two regularization distances, $h=$ \SI{2.5}{mm} or \SI{10}{\micro m}, are shown as filled or open data points, respectively.}
		\label{Fig_CylinderIntegral}
	\end{figure}

	By integrating over a larger cylinder height, $H$, the volume-averaged JNN decreases as a result of averaging over uncorrelated noise at relatively larger distances. Results of $\left\langle \bar{\bar{B}}_z(R, H, h)^2 \right\rangle^{1/2}$ with various $H$ and $R$ are displayed in Fig.\,\ref{Fig_CylinderIntegral}. Again, filled and open points were computed with $h=$ \SI{2.5}{mm} and \SI{10}{\micro m}, respectively. 		
	
	From Figs.\,\ref{Fig_DiskVsCylinder} and \ref{Fig_CylinderIntegral}, one can see that the larger the cylinder volume ($R$ or $H$), the smaller the influence of the regularization distance, $h$.

	\subsection{Finite-element method with discrete dipoles} \label{Sec_FEM}
	To estimate the JNN originating from the electrodes, instead of infinite slabs, conductors of finite size need to be considered. Lee and Romalis\,\cite{Lee2008} calculated magnetic noise from conducting objects of simple geometries. The JNN calculation for a thin circular planar conductor is shown in Tab.\,VI of Ref.\,\cite{Lee2008}, which accords with the geometry of the electrodes and was used in our study. To estimate an upper limit of the average-field difference observed by UCN and HgM in the presence of JNN, it is sufficient to apply the static approximation, where only a \textit{white-noise} spectrum at the limit of $f\rightarrow 0$ should be considered, as shown in Fig.\,\ref{Fig_Bz}. The $z$ component of the ASD measured at a normal distance $d$ generated by a thin film of radius $R$, thickness $a$ and  conductivity $\sigma$ at a temperature $T$ is\,\cite{Lee2008}\,\footnote{Compare to the original expression shown in Ref.\,\cite{Lee2008}, where variables $a$, distance, $t$, thickness, and $r$, radius, were replaced with $d$, $a$ and $R$ in our paper to avoid confusion and keep consistency with other variables used in the remainder of the article.} 
	\begin{equation} \label{Eq_Bthin}
		\frac{1}{\sqrt{8 \pi}} \frac{\mu_{0}\sqrt{k_{{\rm B}}T\sigma a}}{d}\frac{1}{1+\frac{d^{2}}{R^{2}}} =: \mathcal{B}_{z}^{{\rm thin}}(d, f\rightarrow 0). 
	\end{equation}
	However, for a calculation of the magnetic field sampled by particles within the chamber, the JNN spectrum shown above, Eq.\,\eqref{Eq_Bthin}, which depends only on the normal distance between the source of noise and the observation point need to be replaced with a time-dependent three-dimensional magnetic-noise source. For this reason, we used a supplementary method by considering a finite number of random magnetic dipoles on the surface of the conductors as noise sources. 
	
	Although the n2EDM apparatus consists of a double-chamber, to study the possible cancellation of field fluctuation deploying a co-magnetometer system, we considered only the field measurements taken place in one precession chamber. In general, only the contribution from the two electrodes defining top and bottom of the relevant chamber needs to be considered, as the effect from metal plates further away is exponentially suppressed. Random time series were generated for each discrete dipole on both electrodes. The superposition of all these time series and the applied field $B_0$ at discrete positions within the volume of the chamber permits to calculate the distinct magnetic field $\boldsymbol{B}(\boldsymbol{r}, t)$ for any time and locations.   
	
	Following the idea of equivalent current dipoles introduced initially in Refs.\,\cite{Varpula1984, Nenonen1996} described in Sec.\,\ref{Sec_AnalyticalDerivation}, we divided the surface of the electrodes into triangular areas. The motivation of using triangular grids instead of common quadrilateral meshing methods will be explained later. For each triangular element, three noise-current sources located at the center of the triangle and oriented along the three Cartesian coordinates representing the normal component of the three directions were created. For a sampling period of $\Delta t$, the magnetic field created by a number of $n_{\rm dip}$ dipoles and measured at position $\boldsymbol{r}$ can be calculated by the discrete Biot-Savart law, 
	\begin{equation} \label{Eq_BiotSavartMod}
		\boldsymbol{B}(\boldsymbol{r}, \Delta t)=\frac{\mu_{0}}{4\pi} \sum_{i=1}^{n_{\rm dip}} \sum_{\alpha=x,y,z} \frac{I_{\alpha, i} (\Delta t) \diff{\boldsymbol{l}} \times (\boldsymbol{r}-\boldsymbol{r_{i}'})}{|\boldsymbol{r}-\boldsymbol{r_{i}'}|^{3}}, 
	\end{equation}
	where $\boldsymbol{r_{i}'}$ and $I_{\alpha,i}(\Delta t)$ are the position and current of an individual dipole in direction $\alpha$, and the unit-length vector $\diff{\boldsymbol{l}}$ was defined to be the average side length of the triangles.  
	
	The white-noise ASD of the thermal current is\,\cite{Varpula1984, Nenonen1996, Lee2008}  
	\begin{equation} \label{Eq_Ispectrum}
		\mathcal{I}=\sqrt{k_{\rm B}T\sigma a},
	\end{equation} 
	having a unit of A/$\sqrt{\rm Hz}$. By using the power spectral density (PSD = ASD$^{2}$), the variance of the dipole-current time series can be calculated as 
	\begin{align} \label{Eq_TimeVarN2Edm}
		\sigma^{2}(I(\Delta t)) &= 2\, \mathcal{I}^2 \Delta f_{\text{\tiny{\rm BW}}} \nonumber \\ 
		&=2 k_{\rm B}T\sigma a \Delta f_{\text{\tiny{\rm BW}}},
	\end{align}
	where $\Delta f_{\text{\tiny{\rm BW}}}$ is the bandwidth, corresponding to $1/(2 \Delta t)$ with $\Delta t$ being the average observation time. The current $I_{\alpha, i} (\Delta t)$ in Eq.\,\eqref{Eq_BiotSavartMod} is a random current drawn from a Gaussian distribution with the defined variance, $\sigma^{2}(I(\Delta t))$.
	
\begin{figure}[h!]
		\centering
		\includegraphics[width=.95\linewidth]{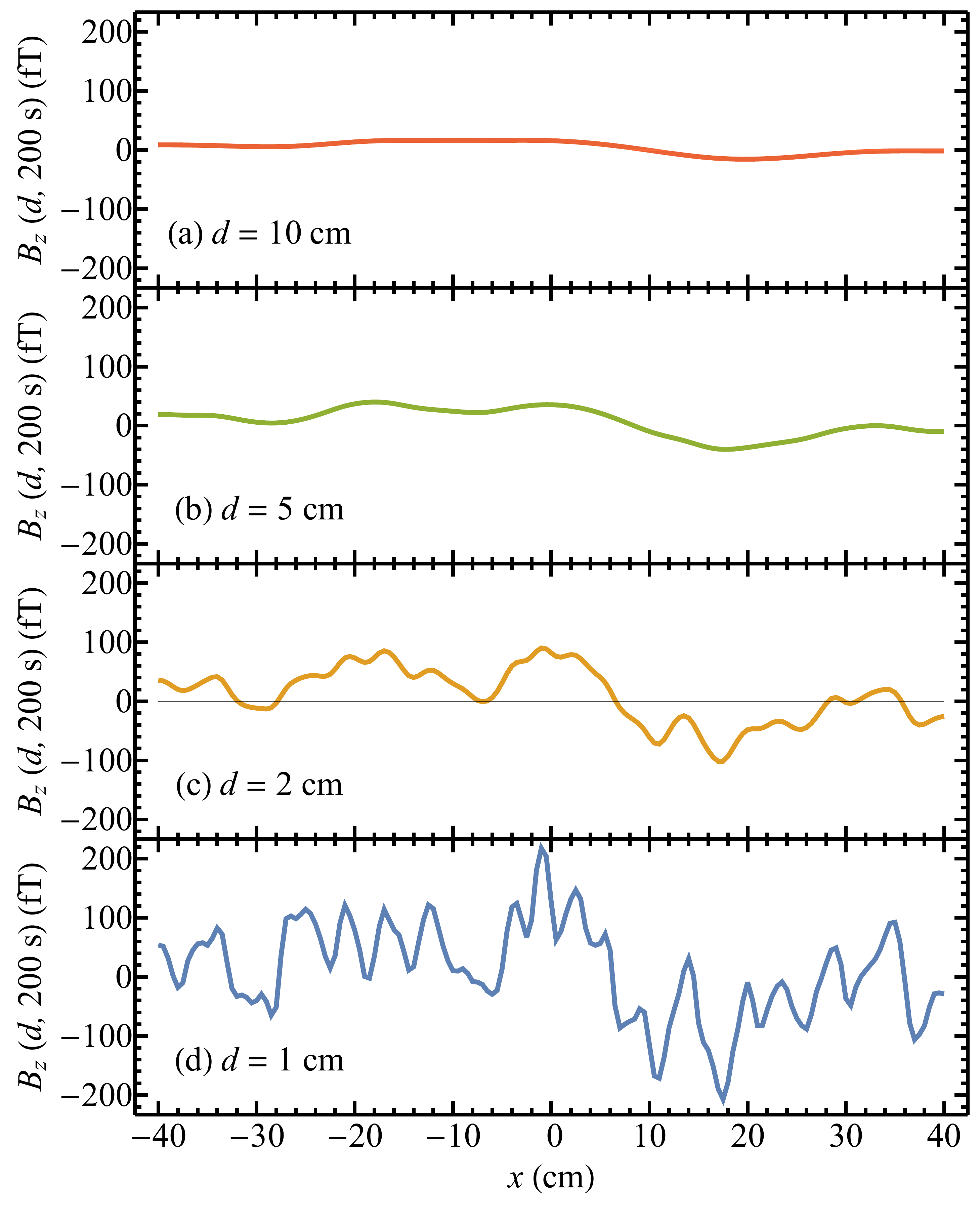}
		\caption{Normal component of time-domain magnetic-field distribution along a horizontal cut line at various distances. This is an example created from one random current-dipole set.}
		\label{Fig_hLineField}
	\end{figure}

	The surface of the aluminum electrode was divided into approximately 1500 finite surface elements, whose average side length was about 28\,mm. 
	The electrode-division layout was optimized to provide a theoretically compatible noise spectrum and to be computationally efficient. A dipole with three time-averaged current components lies at the center of each surface element. The current components were randomly created from a Gaussian distribution with a defined variance using $\Delta t = 200$~s, which is the free-spin-precession period of one measurement. Figure \ref{Fig_hLineField} shows the time-domain magnetic-field distribution along a horizontal cut line, $x = -40\dots 40$~cm within the diameter of the chamber, with one random current-dipole set, where (a)-(d) indicate field distributions at various distances from the electrode. The shorter the distance to the source, the larger the amplitude of the field. Note that also the fluctuation of the field is larger in close vicinity to the source. This validates the argument in the beginning of this section, that a normal-distance dependent JNN spectrum, Eq.\,\eqref{Eq_Bthin}, is inadequate for the purpose of calculating the impact of JNN on the sensitivity of field-sensing spin-$\frac{1}{2}$ particles, since the spatial correlation between adjacent observation points is not considered in the spectral-density formulation.  
	
	This finite-element method was used to calculate the time-and-volume-averaged magnetic field observed by field-sensing particles within the chamber over one measurement, in the presence of JNN\@. Monte Carlo simulations matching experimental results\,\cite{Zsigmond2018} show that the whole volume of the chamber is sampled isotropically during one 200 s measurement in the case of a large number of particles; therefore, it is sufficient to divide the chamber into equally-sized finite volumes, and calculate the magnetic fields observed at the center of each of these rectangular cuboids. A good balance between numerical accuracy and computational efficiency was reached with a size of $10\!\times\!10\!\times 5 $~mm$^{3}$ for these voxels. The reason for using a smaller vertical dimension was due to the fact that, according to the noise spectrum, JNN is normal-distance dependent; hence, a transverse separation between two observation points has fewer impacts than a vertical separation. This was confirmed by using voxels with a size of $5\!\times\!5\!\times 5 $~mm$^{3}$ whose result was comparable to that calculated with a $\SI{10}{mm}$ transverse dimension. In addition, the volume that was divided has a six times larger diameter than height; to partition the chamber in both transverse and vertical directions into numbers of units with the same order of magnitude, we decided to use a voxel with a smaller vertical size. Voxels with a size of  $10\!\times\!10\!\times 2 $~mm$^{3}$ were also studied which gave a negligible difference. As a result, for a better computational efficiency, a voxel size of $10\!\times\!10\!\times 5 $~mm$^{3}$ was selected as the optimal size for chamber partition and used for the results presented below. 
	
	The reason for using triangular meshes for the conductor is to avoid aligned patterns between the conductor grids and the voxels within the chamber volume. In a preliminary study, we found that an alignment between the two meshing patterns could result in artifacts that computed extremely large magnetic-field values due to minimum distances between the noise sources and the observation points. This should be avoided and was resolved by implementing different meshing geometries for the conductors and the chamber where alignments could be well reduced.

	\subsection{Comparison between the analytical derivation and the finite-element method} \label{Sec_CfAnalyticalFEM}
	
\begin{figure}[h!]
		\centering
		\includegraphics[width=.95\linewidth]{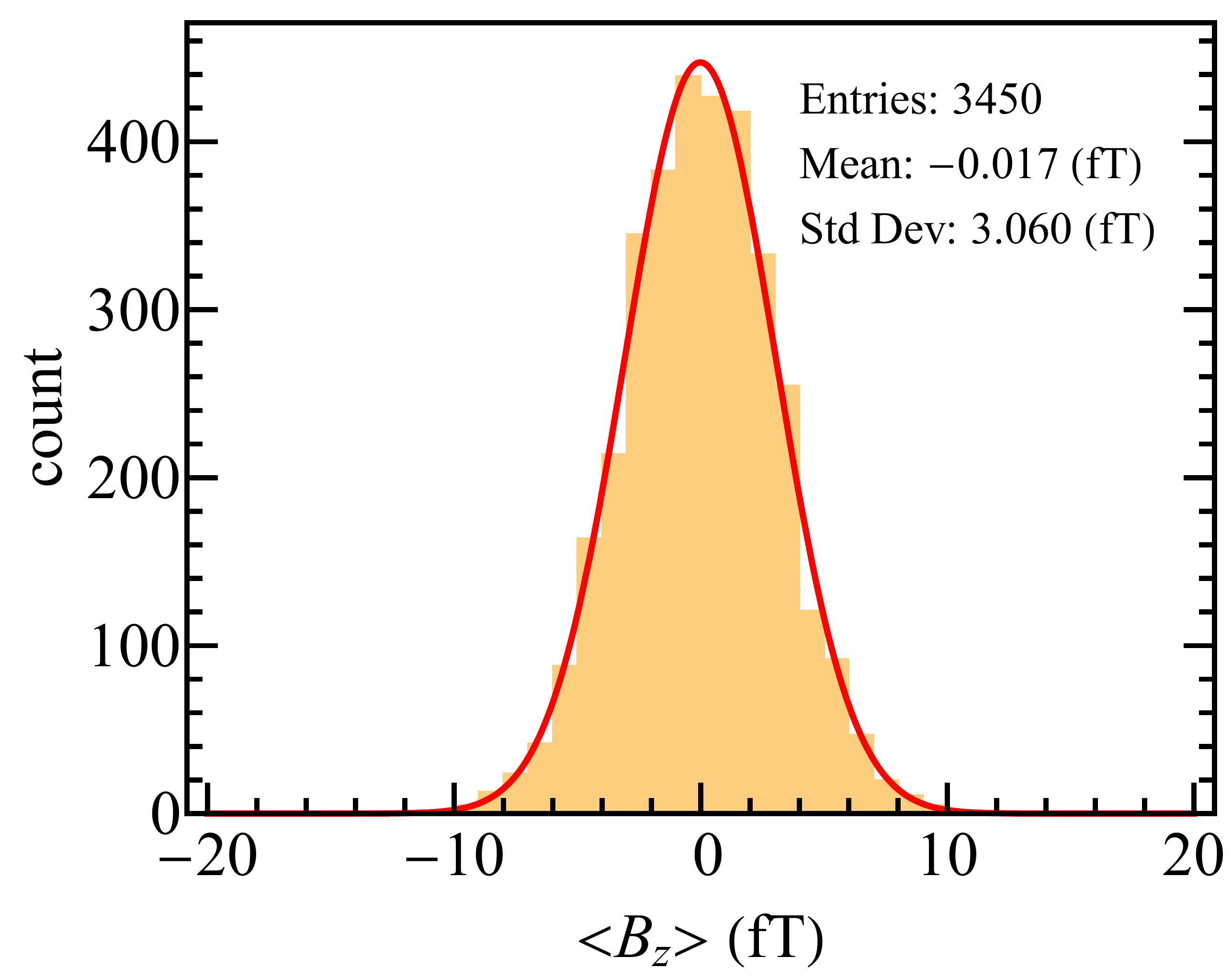}
		\caption{Volume average of normal JNN component $\left\langle B_z \right\rangle$ over half of the n2EDM chamber calculated with the numerical finite-element method.}
		\label{Fig_BzHist}
	\end{figure}

	With the finite-element method, the magnetic noise averaged over a disk or a cylinder can be easily estimated. With the optimal voxel height of \SI{5}{mm}, the n2EDM precession chamber was divided into 24 layers each consists of 5024 pieces. 
	First, we calculated the average field over different numbers of adjacent voxels on the same layer, corresponding to radii ranging from $10\dots35$\,mm, w.r.t.\ a {\it central} piece, where 100 random central pieces were selected. Results from four different layers, with distances of $7.5\dots22.5$\,mm, are shown in distinct colors and shapes in Fig.\,\ref{Fig_DiskIntegral}. 
	The error bar on each point is the standard deviation of these 100 randomly chosen central pieces. We confirmed that the surface-averaged magnetic-field noise computed with the finite-element method is in good agreement with both the analytical derivation and the calculation performed in Ref.\,\cite{Nenonen1996}. 
	Next, to calculate a cylinder average with the finite-element method compatible with the analytical description, we considered only half of the precession chamber and one electrode. The half-chamber volume average of normal magnetic field generated by magnetic dipoles on this electrode, $\left\langle B_z \right\rangle$, was computed and shown in Fig.\,\ref{Fig_BzHist}. Each entry in the histogram is the result of one simulated cycle. For one finite-element calculation, i.e.\ one simulated cycle, approximately 1500 dipoles were created on the conductor using three random noise currents at a bandwidth of $\Delta f_{\text{\tiny{\rm BW}}} = 1/(2\!\times\!200\,{\rm s})$. A total of more than 3000 random configurations were generated to accumulate statistics. The standard deviation of these random solutions is $\sigma_{Bz} = \SI[separate-uncertainty]{3.060(37)}{fT}$. The uncertainty is the standard error of $\sigma_{Bz}$ estimated theoretically with $SE(\sigma_{Bz}) = \sigma_{Bz}/\sqrt{2S-2}$, where $S=3450$ is the number of simulations. On the other hand, using the analytical formula of volume variance, Eq.\,\eqref{Eq_VolAvgNoise}, and replacing the infinite conductor with a finite conductor of $R=\SI{40}{cm}$, the standard deviation of the volume average with twenty random solutions for the Monte Carlo integration is  
	\begin{equation}	
		\left\langle \bar{\bar{B}}_z(40\,{\rm cm}, 6\,{\rm cm}, 2.5\,{\rm mm})^2 \right\rangle^{1/2} = 2.805 \pm 0.005\,{\rm fT},
	\end{equation}
	where the error is the statistical error of the Monte Carlo sample. The results from the two methods agree within half a femtotesla. The small deviation is negligible for our purpose and confirms the use of voxels for volume-average calculation with the finite-element method. 	 	

	\section{Effects on the \texorpdfstring{\lowercase{n}2EDM}{n2EDM} experiment}	
	\subsection{Magnetic fields observed by UCN and HgM}	
	Due to the difference in the velocity spectrum and the Larmor precession frequency, UCN and HgM sample the volume differently under a nominal \SI{1}{\micro T} $B_{0}$ field. Much faster thermal $^{199}$Hg atoms fall into the non-adiabatic regime. The spins precess under a vectorial volume average of the field; hence, the average magnetic field observed by $^{199}$Hg atoms is calculated as 
	\begin{equation} \label{Eq_BsampledHg}
		\begin{split}
			\left\langle B_\text{\tiny{Hg}} \right\rangle &= \left| \left\langle \boldsymbol{B} \right\rangle \right| \\ 
			&= \sqrt{\left\langle B_x \right\rangle ^{2} + \left\langle B_y \right\rangle ^{2} + \left\langle B_0 + B_z \right\rangle ^{2}}. 
		\end{split} 
	\end{equation}
	By contrast, due to the much smaller velocity and larger precession frequency, UCN sample the volume in the adiabatic regime, such that their spins precess under the volume average of the modulus of the field. In addition, taking into account the negative center-of-mass offset $\left\langle z \right\rangle$ of the ensemble of UCN, the average field sampled by UCN is   
	\begin{equation} \label{Eq_BsampledUcn}
		\begin{split}
			\left\langle B_\text{\tiny{UCN}} \right\rangle &= \left\langle \left| \boldsymbol{B} \right|  \rho_{\text{{\tiny UCN}}}(z) \right\rangle \\ 
			&= \left\langle \sqrt{B_x\!^{2} + B_y\! ^{2} + \left( B_0 + B_z \right) ^{2}} \rho_{\text{{\tiny UCN}}}(z) \right\rangle,
		\end{split}
	\end{equation}
	where
	\begin{equation} \label{Eq_ComOffset}
		\rho_{\text{{\tiny UCN}}}(z) =\frac{1}{H}\left( 1 + \frac{12 \left\langle z \right\rangle }{H^{2}} z\right)
	\end{equation}
	is the normalized vertical UCN density function.
	
\begin{figure}
		\centering
		\begin{subfigure}[h!]{.45\textwidth}
			\includegraphics[width=\linewidth]{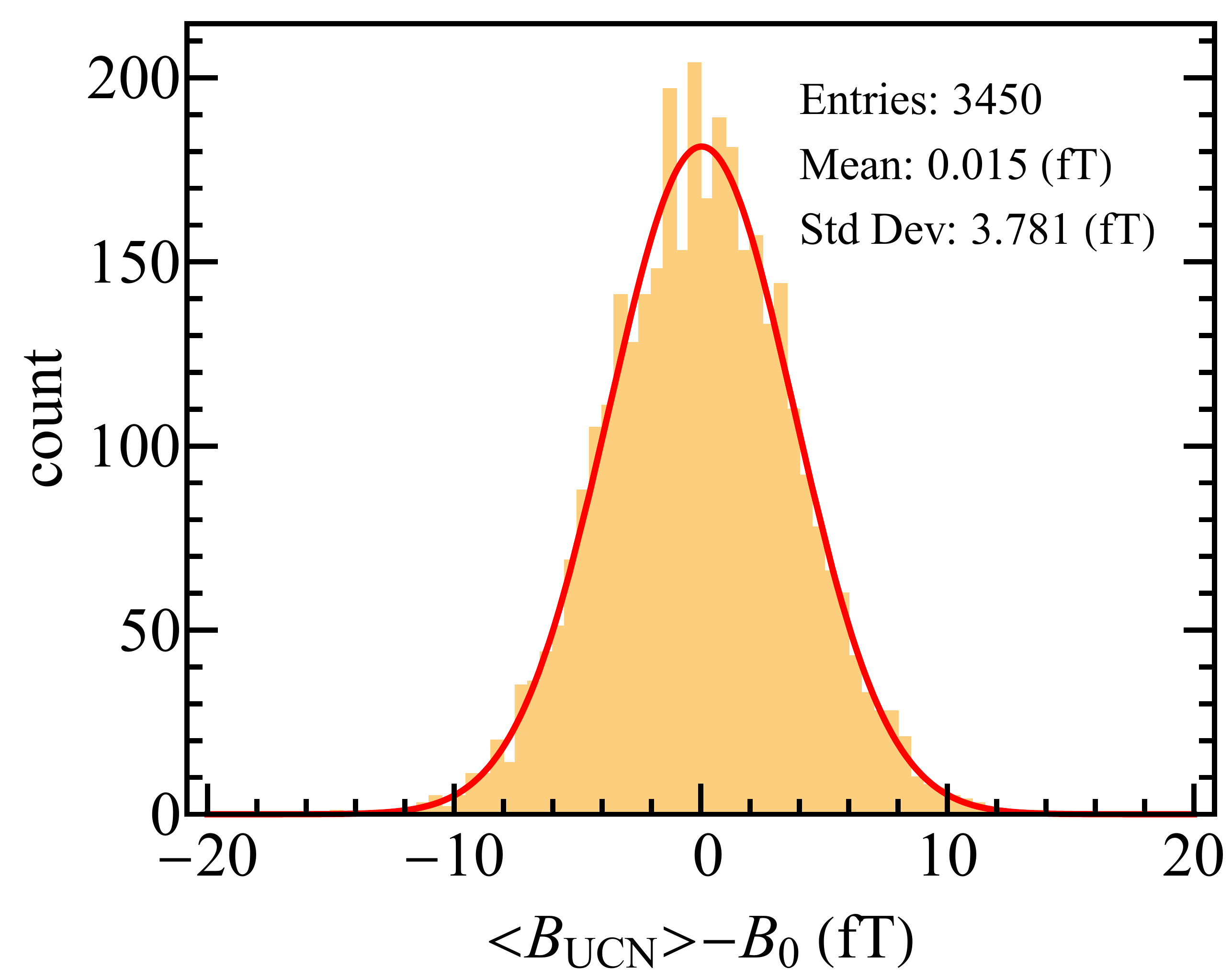}
			\caption{} \label{SubFig_HistUcn}
		\end{subfigure}
		\begin{subfigure}[h!]{.45\textwidth}
			\includegraphics[width=\linewidth]{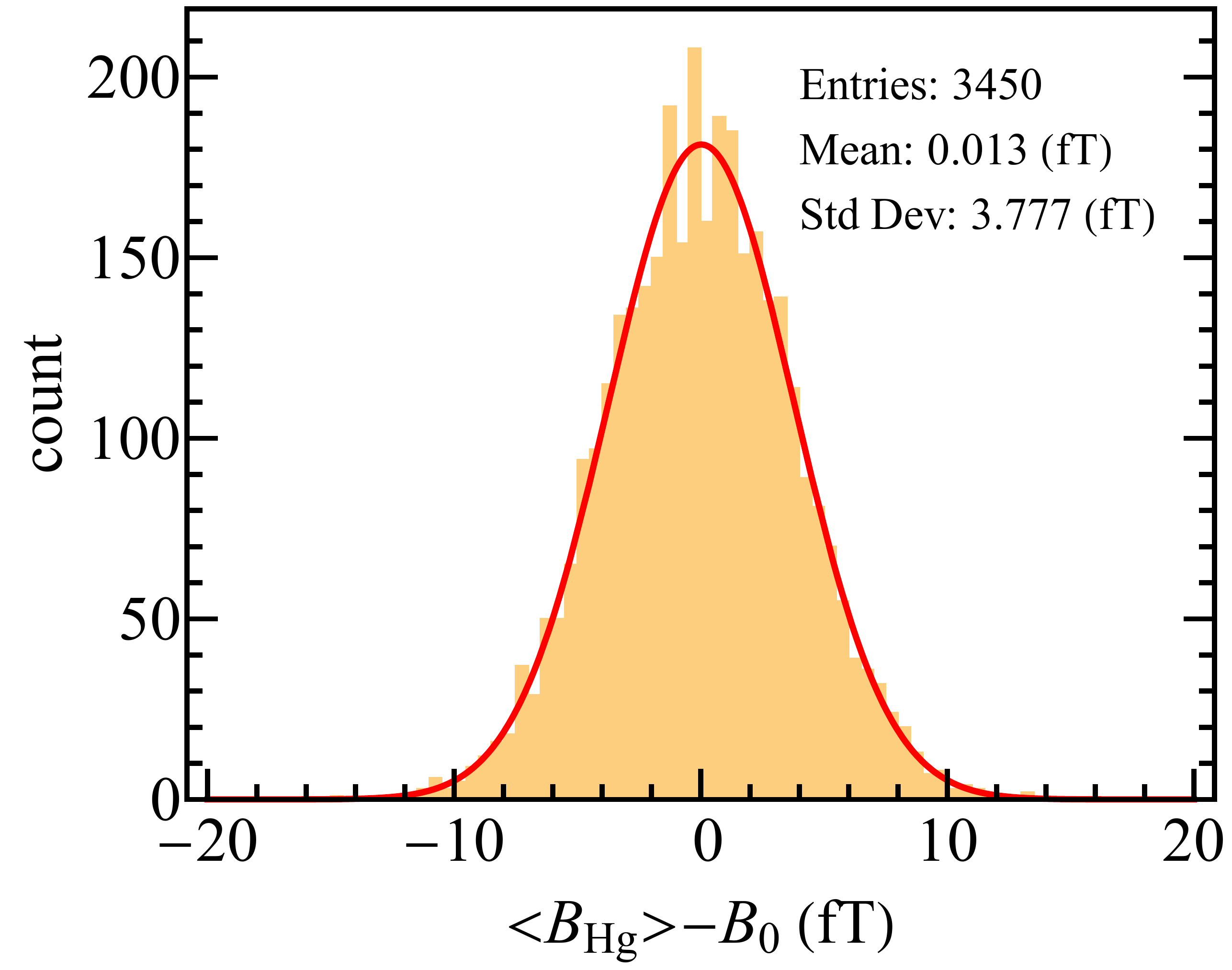}
			\caption{} \label{SubFig_HistHg}
		\end{subfigure}
		\caption{Deviations of time-and-volume-averaged field to the nominal constant $B_{0}$ magnetic field sampled by (a)~UCN and (b)~$^{199}$Hg atoms.} 
		\label{Fig_BsepHist}		  
	\end{figure}

	To estimate the time-and-volume average of the magnetic fields over one precession chamber sandwiched between two electrodes, the finite-element method was employed. Figure~\ref{Fig_BsepHist} shows the average magnetic fields observed by UCN and HgM over one simulated cycle, calculated with $\left\langle z \right\rangle =-4.1$\,mm for Eq.\,\eqref{Eq_ComOffset}. This offset value was obtained using a Monte Carlo simulation\,\cite{Zsigmond2018} and is in agreement with the offset obtained in Ref.\,\cite{Abel2020}. Each entry in the histogram in Fig.\,\ref{Fig_BsepHist} is the result of one $\Delta t = 200$\,s time average simulated with approximately 1500 dipoles created on both electrodes. The average magnetic fields were computed with Eqs.\,\eqref{Eq_BsampledHg} and \eqref{Eq_BsampledUcn}. Histograms for UCN and $^{199}$Hg atoms are shown in Figs.\,\ref{SubFig_HistUcn} and \ref{SubFig_HistHg}, respectively. The standard deviation of these distributions, $\sigma \left(\left\langle B_\text{\tiny{UCN}} \right\rangle\right) = \SI[separate-uncertainty]{3.781(46)}{fT} \approx \sigma \left(\left\langle B_\text{\tiny{Hg}} \right\rangle\right) = \SI[separate-uncertainty]{3.777(46)}{fT}$, are comparable within the statistical error, confirming the naive hypothesis that they similarly sense the effects from JNN\@. In addition, this is an order of magnitude lower than the sensitivity requirement, 30 fT per measurement cycle, for the HgM in the n2EDM experiment\,\cite{Ayres2021}, indicating that the performance of HgM will not be limited by JNN\@.   
	
\begin{figure}[t!]
		\centering
		\includegraphics[width=.95\linewidth]{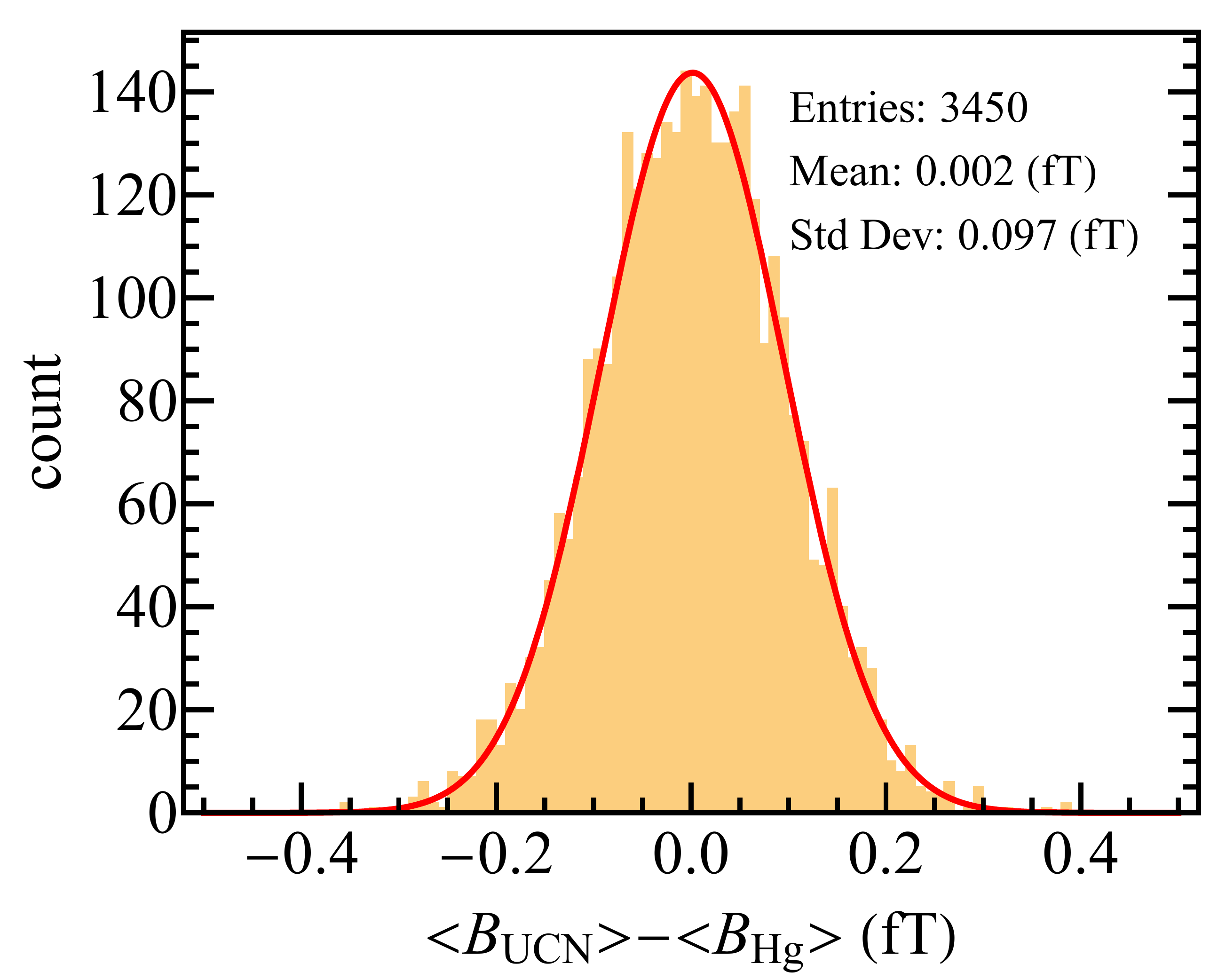}	
		\caption{Difference between the average fields sampled by UCN and mercury ensembles, $\left\langle B_\text{\tiny{UCN}} \right\rangle - \left\langle B_\text{\tiny{Hg}} \right\rangle$.}
		\label{Fig_BdiffHist}	
	\end{figure}

	Next, we studied the influence of JNN on the ratio of precession frequencies of the two spin-$\frac{1}{2}$ ensembles by looking at the difference of the average magnetic fields (see Fig.\,\ref{Fig_BdiffHist}).
	The standard deviation of the differences of average magnetic fields is $\sigma \left(\left\langle B_\text{\tiny{UCN}} \right\rangle - \left\langle B_\text{\tiny{Hg}} \right\rangle\right) \sim 0.1\,{\rm fT}$. The sensitivity of an nEDM measurement depends on the uncertainty of the magnetic-field measurement. By using a mercury co-magnetometer, the effect from JNN is reduced to $\sigma_{\text{\tiny{\rm JNN}}} \sim 0.1\,{\rm fT}$ per measurement, and induces an uncertainty on the neutron EDM of 
	\begin{equation} \label{Eq_SigmaDn}
		\sigma_{d_{\rm n}} = \frac{\hbar}{2E} \gamma_{\textrm{n}} \sigma_{\text{\tiny{\rm JNN}}} = 4 \times 10^{-28}~e\!\cdot\!{\rm cm},
	\end{equation}
	assuming an electric field $E=15$~\si{kV/cm} and $\gamma_{\rm n}/2\pi=29.16$~\si{MHz/T} is the gyro-magnetic ratio of the neutron. The experiment will consist of a total of $M$ \SI{200}{s} long measurement cycles to improve the statistical sensitivity. Note that the uncertainty on $\sigma_{d_{\rm n}}$ calculated for one cycle in Eq.\,\eqref{Eq_SigmaDn} scales statistically with $M^{-1/2}$. Results shown in Figs.\,\ref{Fig_BsepHist} and \ref{Fig_BdiffHist} had been crosschecked with another 2000 random configurations which showed similar results, confirming the negligibility of the statistical error arising from the sampling size.  
	
	\subsection{Magnetic field measured by CsM}		
	The design of the n2EDM experiment\,\cite{Ayres2021} deploys more than 100 cesium magnetometers (CsM) being installed above and below the precession-chamber stack in order to provide essential information about the homogeneity and stability of the magnetic field. They are arranged radially in groups of four on vertical modules. Each CsM contains a glass bulb filled with saturated vapor pressure of $^{133}$Cs atoms. They are operated as Bell-Bloom type\,\cite{Bell1961} magnetometers. Tensor-polarization (alignment) is created by amplitude-modulated linearly-polarized laser light that traverses the bulb, at a frequency roughly matched with the $^{133}$Cs Larmor precession frequency, similar as in Refs.\,\cite{Grujic2015, Afach2015d}. Once the atomic vapor is spin aligned, the light intensity is reduced and kept constant. As the spin-polarized atoms precess under the influence of $\boldsymbol{B}$ with a frequency proportional to the magnitude of the magnetic field, the intensity of transmitted light is periodically modulated by precessing atoms and detected by a photodiode. 
	
	Consider a CsM with a radius of 1.5\,cm placed above the top-most electrode. Polarized $^{133}$Cs atoms at different locations within the bulb are exposed to magnetic noise from the electrode which decreases with distance according to Eq.\,\eqref{Eq_RmsLsdZ}. The finite-element method introduced in Sec.\,\ref{Sec_FEM} was employed to calculate the average magnetic field measured by $^{133}$Cs atoms in the presence of JNN\@. For a CsM bulb, the measurement time of the magnetic field is $\delta t = 70$\,ms, which is roughly two times the spin-coherence time of $^{133}$Cs atoms. The skin depth at \SI{14}{Hz} is 2.2\,cm which is outside of the thickness range in which the static approximation is valid. Nonetheless, the static approximation can be used to obtain an upper limit for the field fluctuation. 
	
\begin{figure}[h!]
		\centering
		\includegraphics[width=.95\linewidth]{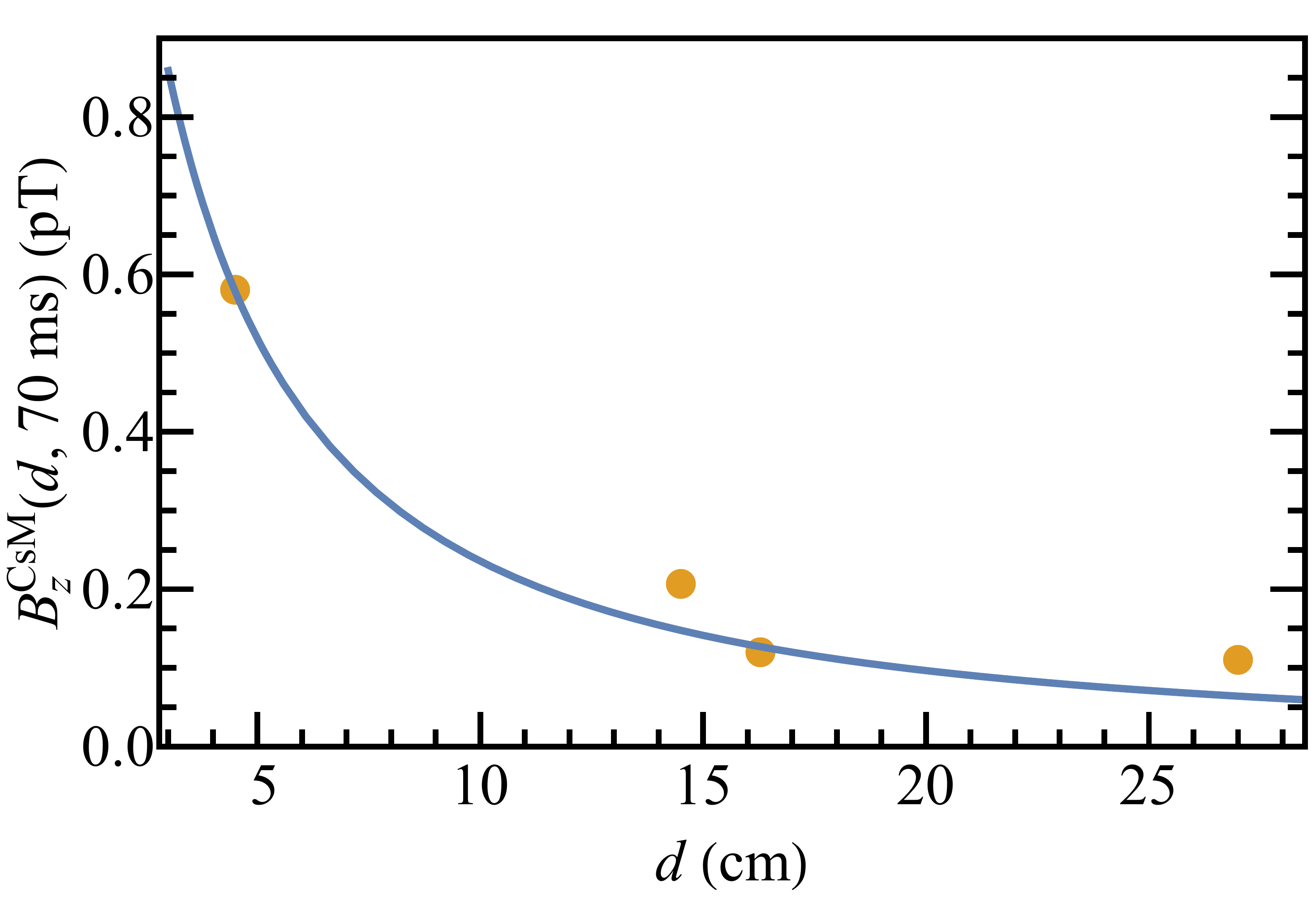}
		\caption{Comparison between RMS normal noise amplitude integrated over 70 ms, calculated from the noise spectrum (blue line), and the average-field noise measured by a CsM estimated with the finite-element method (orange points). The statistical errors on the results calculated with the finite-element method are smaller than the marker size.} 
		\label{Fig_JnnCsM}
	\end{figure}

	In this case, only the closest electrode which was relevant to a specific CsM was considered. Similarly, the noise source was represented by a number of dipoles lying on the surface of the electrode, each with three random noise currents at the bandwidth of $\Delta f_{\text{\tiny{\rm BW}}} = 1/(2\!\times\!70\,{\rm ms})$. The bulb was divided into about 14000 voxels of size 1\,mm$^{3}$, much smaller than the voxel size used for chamber division due to the orders of magnitude smaller volume. The $^{133}$Cs atoms sense the field in the same  way as the $^{199}$Hg atoms in the precession chamber. Hence, the average magnetic field observed by a CsM was calculated by averaging over the fields in all voxels using Eq.\,\eqref{Eq_BsampledHg}. More than 3000 random dipole sets were simulated. The average magnetic fields from JNN for four CsM bulbs placed on one module with different distances to the electrode were simulated. The corresponding standard deviations of the time-and-volume-averaged fields at these positions are shown as orange points in Fig.\,\ref{Fig_JnnCsM} whose statistical errors are three orders of magnitude smaller.  
	
	In a perfectly spherical CsM bulb, the $^{133}$Cs atoms are uniformly distributed over the volume. Due to the fast movement of $^{133}$Cs atoms, the average magnetic field over the sphere is sampled homogeneously and its value is equal to the field measured at the center based on the mean-value theorem\,\cite{Griffiths2017,Jackson1998}, assuming all sources are outside the sphere. The RMS magnetic noise can be estimated by the noise observed at the center of the bulb within a time span $\delta t$. At a distance $d$ measured to the center of the bulb, the RMS magnetic noise is
	\begin{equation}
		B^{\rm CsM}_{i}(d, \delta t) = 
		\left\lbrace \int_{0}^{\frac{1}{2 \delta t}} \mathcal{B}_{i}\left(d,f \right)^{2} \diff{f} \right\rbrace ^{1/2}, 
		\label{Eq_BcsRms}
	\end{equation}
	with $i$ being $x, y$ or $z$. In the presence of an applied $B_{0}\!\parallel\!B_z$ field of about \SI{1}{\micro T}, the lateral components $B_x,B_y \ll B_0$ of JNN are quadratically suppressed, hence negligible. For this reason, we only take the vertical component into account. The normal RMS magnetic noise estimated at the center of the CsM, $B^{\rm CsM}_{z}(d, \delta t)$, as a function of distance, is also displayed in Fig\,\ref{Fig_JnnCsM}.
	
	In the figure, both methods deliver similar results with small differences which can be understood by the following explanations. Equation\,\eqref{Eq_BcsRms} is the frequency-bandwidth integrated RMS noise generated by an infinite slab, whereas the finite-element method took a finite size of the electrode and only the low-frequency noise was considered. Therefore, the results calculated from the finite-element method will in principle be larger due to the use of static approximation which is true for three of the cases. As for the result calculated at $d = 16.3$\,cm, the finite-element method computed a smaller value.
	For this specific case, the CsM is placed at $R = \SI{55}{cm}$, which is larger than the electrode radius; hence, the effect of noise from the electrode will be smaller compared to the theoretical calculation which used an infinite conductor. In general, this method provides a sufficiently precise estimation of the impact of JNN on the measurements by the CsM.   
	
	The sensitivity goal for n2EDM translates to a maximum RMS noise of 2.7\,pT in 70\,ms for the CsM \cite{Ayres2021}. The upper limits of the noise for CsM at various distances all lie below the sensitivity limit. In addition, for an nEDM-measurement cycle of 200\,s $\gg$ 70\,ms, the magnetic noise seen by a CsM will be averaged out to a much lower value; hence, we confirm that JNN from the electrodes is negligible for the design and placement of all CsM within the experiment.

	\section{Conclusion} 
	This paper reports on a finite-element study of Johnson-Nyquist noise (JNN) originating from the bulk metal electrodes in the n2EDM experiment being constructed by the nEDM collaboration at PSI\@. In the first part, we revisited the theoretical noise spectra\,\cite{Varpula1984, Nenonen1996}, and compared them to the measurements on a thin aluminum sheet using a superconducting quantum interference device (SQUID).  
	
	Next, we derived for a given frequency bandwidth expressions for the root-mean-square normal noise amplitudes of averages over a two-dimensional disk and a cylinder of finite volume. These are important in understanding the spatial correlation of JNN and are necessary for practical purposes. Numerical results from the analytical derivation were computed with the method of Monte Carlo integration and demonstrate good agreement with the calculation performed in the literature\,\cite{Nenonen1996}. 
	
	Using a discretization of the electrodes into a finite number of magnetic white-noise dipoles, we calculated temporal and spatial magnetic fields generated by JNN\@. By averaging these magnetic fields over time and volume, we obtained the mean magnetic field sensed by precessing ultracold neutrons (UCN) and mercury ($^{199}$Hg) atoms. The standard deviation of more than 3000 randomly produced configurations for UCN and mercury is approximately 3.8~fT, which we consider as small enough for next-generation neutron electric-dipole-moment (nEDM) searches. With the same method, we found that for the cesium ($^{133}$Cs) vapor magnetometers, the maximum RMS noise observed within a measurement time of 70~ms is approximately 0.6~pT, which lies below the sensitivity goal of 2.7~pT for n2EDM. Thus, we confirm that the precision of the cesium magnetometers will not be constrained by JNN from the aluminum electrodes. 
	
	Additionally, by computing the average-field difference observed by UCN and mercury, we found that the noise is sensed highly correlated and mostly cancels out by using a co-magnetometer to normalize the UCN measurements. That is, the impact of JNN is negligible for nEDM searches down to a sensitivity of $4\,\times\,10^{-28}\,e\cdot{\rm cm}$ for a single 200\,s measurement. Assuming a projected experiment of 500 days with $\sim$ 280 cycles per day, this results in a factor of 374 smaller limit, which is sufficiently small for our planned nEDM search using a co-magnetometer concept.
	
	\section{Acknowledgments}
	We would like to thank A.~Crivellin and M.~Spira for helpful discussions. We are grateful for the technical support from P.~H\"ommen and R.~K\"orber with the material measurements in BMSR-2, PTB, Berlin. The material measurements inside BMSR-2 were supported by the Core Facility ``Metrology of Ultra-Low Magnetic Fields'' at PTB funded by Deutsche Forshungsgemeinschaft (DFG) through funding codes: DFG KO 5321/3-1 and TR408/11-1. The swiss members acknowledge the financial support from the Swiss National Science Foundation through projects 157079, 163413, 169596, 188700 (all PSI), 181996   (Bern),   172639  (ETH), and  FLARE20FL21-186179. This work has also been supported by the Cluster of Excellence ``Precision Physics, Fundamental Interactions, and Structure of Matter'' (PRISMA + EXC2118/1) funded by DFG within the German Excellence Strategy (Project ID 39083149) from Johannes Gutenberg University Mainz. This work is also supported by Sigma Xi grants \#\,G2017100190747806 and \#\,G2019100190747806, and by the award of the Swiss Government Excellence Scholarships (SERI-FCS) \#\,2015.0594. The group from Jagellionian University Cracow acknowledges the support from National Science Centre, Poland, through grants No.\,2018/30/M/ST2/00319, and No.\,2020/37/B/ST2/02349. The group from University of Leuven acknowledges supports from the Fund for Scientific Research Flanders (FWO).
	
	\bibliography{C:/Users/chiu_p/polybox/UCN/Reports_Presentations/PaperJnn/bib_JnnPaper}	

\begin{thebibliography}{48}%
\makeatletter
\providecommand \@ifxundefined [1]{%
 \@ifx{#1\undefined}
}%
\providecommand \@ifnum [1]{%
 \ifnum #1\expandafter \@firstoftwo
 \else \expandafter \@secondoftwo
 \fi
}%
\providecommand \@ifx [1]{%
 \ifx #1\expandafter \@firstoftwo
 \else \expandafter \@secondoftwo
 \fi
}%
\providecommand \natexlab [1]{#1}%
\providecommand \enquote  [1]{``#1''}%
\providecommand \bibnamefont  [1]{#1}%
\providecommand \bibfnamefont [1]{#1}%
\providecommand \citenamefont [1]{#1}%
\providecommand \href@noop [0]{\@secondoftwo}%
\providecommand \href [0]{\begingroup \@sanitize@url \@href}%
\providecommand \@href[1]{\@@startlink{#1}\@@href}%
\providecommand \@@href[1]{\endgroup#1\@@endlink}%
\providecommand \@sanitize@url [0]{\catcode `\\12\catcode `\$12\catcode
  `\&12\catcode `\#12\catcode `\^12\catcode `\_12\catcode `\%12\relax}%
\providecommand \@@startlink[1]{}%
\providecommand \@@endlink[0]{}%
\providecommand \url  [0]{\begingroup\@sanitize@url \@url }%
\providecommand \@url [1]{\endgroup\@href {#1}{\urlprefix }}%
\providecommand \urlprefix  [0]{URL }%
\providecommand \Eprint [0]{\href }%
\providecommand \doibase [0]{http://dx.doi.org/}%
\providecommand \selectlanguage [0]{\@gobble}%
\providecommand \bibinfo  [0]{\@secondoftwo}%
\providecommand \bibfield  [0]{\@secondoftwo}%
\providecommand \translation [1]{[#1]}%
\providecommand \BibitemOpen [0]{}%
\providecommand \bibitemStop [0]{}%
\providecommand \bibitemNoStop [0]{.\EOS\space}%
\providecommand \EOS [0]{\spacefactor3000\relax}%
\providecommand \BibitemShut  [1]{\csname bibitem#1\endcsname}%
\let\auto@bib@innerbib\@empty
\bibitem [{\citenamefont {Smith}\ \emph {et~al.}(1957)\citenamefont {Smith},
  \citenamefont {Purcell},\ and\ \citenamefont {Ramsey}}]{Smith1957}%
  \BibitemOpen
  \bibfield  {author} {\bibinfo {author} {\bibfnamefont {J.~H.}\ \bibnamefont
  {Smith}}, \bibinfo {author} {\bibfnamefont {E.~M.}\ \bibnamefont {Purcell}},
  \ and\ \bibinfo {author} {\bibfnamefont {N.~F.}\ \bibnamefont {Ramsey}},\
  }\href {\doibase 10.1103/physrev.108.120} {\bibfield  {journal} {\bibinfo
  {journal} {Phys. Rev.}\ }\textbf {\bibinfo {volume} {108}},\ \bibinfo {pages}
  {120} (\bibinfo {year} {1957})}\BibitemShut {NoStop}%
\bibitem [{\citenamefont {Ramsey}(1950)}]{Ramsey1950}%
  \BibitemOpen
  \bibfield  {author} {\bibinfo {author} {\bibfnamefont {N.~F.}\ \bibnamefont
  {Ramsey}},\ }\href {\doibase 10.1103/physrev.78.695} {\bibfield  {journal}
  {\bibinfo  {journal} {Phys. Rev.}\ }\textbf {\bibinfo {volume} {78}},\
  \bibinfo {pages} {695} (\bibinfo {year} {1950})}\BibitemShut {NoStop}%
\bibitem [{\citenamefont {Purcell}\ and\ \citenamefont
  {Ramsey}(1950)}]{Purcell1950}%
  \BibitemOpen
  \bibfield  {author} {\bibinfo {author} {\bibfnamefont {E.~M.}\ \bibnamefont
  {Purcell}}\ and\ \bibinfo {author} {\bibfnamefont {N.~F.}\ \bibnamefont
  {Ramsey}},\ }\href {\doibase 10.1103/physrev.78.807} {\bibfield  {journal}
  {\bibinfo  {journal} {Phys. Rev.}\ }\textbf {\bibinfo {volume} {78}},\
  \bibinfo {pages} {807} (\bibinfo {year} {1950})}\BibitemShut {NoStop}%
\bibitem [{\citenamefont {Johnson}(1928)}]{Johnson1928}%
  \BibitemOpen
  \bibfield  {author} {\bibinfo {author} {\bibfnamefont {J.~B.}\ \bibnamefont
  {Johnson}},\ }\href {\doibase 10.1103/physrev.32.97} {\bibfield  {journal}
  {\bibinfo  {journal} {Phys. Rev.}\ }\textbf {\bibinfo {volume} {32}},\
  \bibinfo {pages} {97} (\bibinfo {year} {1928})}\BibitemShut {NoStop}%
\bibitem [{\citenamefont {Nyquist}(1928)}]{Nyquist1928}%
  \BibitemOpen
  \bibfield  {author} {\bibinfo {author} {\bibfnamefont {H.}~\bibnamefont
  {Nyquist}},\ }\href {\doibase 10.1103/physrev.32.110} {\bibfield  {journal}
  {\bibinfo  {journal} {Phys. Rev.}\ }\textbf {\bibinfo {volume} {32}},\
  \bibinfo {pages} {110} (\bibinfo {year} {1928})}\BibitemShut {NoStop}%
\bibitem [{\citenamefont {Varpula}\ and\ \citenamefont
  {Poutanen}(1984)}]{Varpula1984}%
  \BibitemOpen
  \bibfield  {author} {\bibinfo {author} {\bibfnamefont {T.}~\bibnamefont
  {Varpula}}\ and\ \bibinfo {author} {\bibfnamefont {T.}~\bibnamefont
  {Poutanen}},\ }\href {\doibase 10.1063/1.332990} {\bibfield  {journal}
  {\bibinfo  {journal} {J. Appl. Phys.}\ }\textbf {\bibinfo {volume} {55}},\
  \bibinfo {pages} {4015} (\bibinfo {year} {1984})}\BibitemShut {NoStop}%
\bibitem [{\citenamefont {Clem}(1987)}]{Clem1987}%
  \BibitemOpen
  \bibfield  {author} {\bibinfo {author} {\bibfnamefont {J.~R.}\ \bibnamefont
  {Clem}},\ }\href {\doibase 10.1109/tmag.1987.1065127} {\bibfield  {journal}
  {\bibinfo  {journal} {{IEEE} Trans. Magn.}\ }\textbf {\bibinfo {volume}
  {23}},\ \bibinfo {pages} {1093} (\bibinfo {year} {1987})}\BibitemShut
  {NoStop}%
\bibitem [{\citenamefont {Nenonen}\ \emph {et~al.}(1996)\citenamefont
  {Nenonen}, \citenamefont {Montonen},\ and\ \citenamefont
  {Katila}}]{Nenonen1996}%
  \BibitemOpen
  \bibfield  {author} {\bibinfo {author} {\bibfnamefont {J.}~\bibnamefont
  {Nenonen}}, \bibinfo {author} {\bibfnamefont {J.}~\bibnamefont {Montonen}}, \
  and\ \bibinfo {author} {\bibfnamefont {T.}~\bibnamefont {Katila}},\ }\href
  {\doibase 10.1063/1.1147514} {\bibfield  {journal} {\bibinfo  {journal} {Rev.
  Sci. Instrum.}\ }\textbf {\bibinfo {volume} {67}},\ \bibinfo {pages} {2397}
  (\bibinfo {year} {1996})}\BibitemShut {NoStop}%
\bibitem [{\citenamefont {K{\"o}rber}(2018)}]{Koerber2018}%
  \BibitemOpen
  \bibfield  {author} {\bibinfo {author} {\bibfnamefont {R.}~\bibnamefont
  {K{\"o}rber}},\ }in\ \href {\doibase 10.1007/978-981-10-5122-7_199} {\emph
  {\bibinfo {booktitle} {{EMBEC} {\&} {NBC} 2017}}},\ Vol.~\bibinfo {volume}
  {65},\ \bibinfo {editor} {edited by\ \bibinfo {editor} {\bibfnamefont
  {H.}~\bibnamefont {Eskola}}, \bibinfo {editor} {\bibfnamefont
  {O.}~\bibnamefont {V{\"a}is{\"a}nen}}, \bibinfo {editor} {\bibfnamefont
  {J.}~\bibnamefont {Viik}}, \ and\ \bibinfo {editor} {\bibfnamefont
  {J.}~\bibnamefont {Hyttinen}}}\ (\bibinfo  {publisher} {Springer,
  Singapore},\ \bibinfo {year} {2018})\ pp.\ \bibinfo {pages}
  {795--798}\BibitemShut {NoStop}%
\bibitem [{\citenamefont {Allred}\ \emph {et~al.}(2002)\citenamefont {Allred},
  \citenamefont {Lyman}, \citenamefont {Kornack},\ and\ \citenamefont
  {Romalis}}]{Allred2002}%
  \BibitemOpen
  \bibfield  {author} {\bibinfo {author} {\bibfnamefont {J.~C.}\ \bibnamefont
  {Allred}}, \bibinfo {author} {\bibfnamefont {R.~N.}\ \bibnamefont {Lyman}},
  \bibinfo {author} {\bibfnamefont {T.~W.}\ \bibnamefont {Kornack}}, \ and\
  \bibinfo {author} {\bibfnamefont {M.~V.}\ \bibnamefont {Romalis}},\ }\href
  {\doibase 10.1103/PhysRevLett.89.130801} {\bibfield  {journal} {\bibinfo
  {journal} {Phys. Rev. Lett.}\ }\textbf {\bibinfo {volume} {89}},\ \bibinfo
  {pages} {130801} (\bibinfo {year} {2002})}\BibitemShut {NoStop}%
\bibitem [{\citenamefont {Jones}\ \emph {et~al.}(2003)\citenamefont {Jones},
  \citenamefont {Vale}, \citenamefont {Sahagun}, \citenamefont {Hall},\ and\
  \citenamefont {Hinds}}]{Jones2003}%
  \BibitemOpen
  \bibfield  {author} {\bibinfo {author} {\bibfnamefont {M.~P.~A.}\
  \bibnamefont {Jones}}, \bibinfo {author} {\bibfnamefont {C.~J.}\ \bibnamefont
  {Vale}}, \bibinfo {author} {\bibfnamefont {D.}~\bibnamefont {Sahagun}},
  \bibinfo {author} {\bibfnamefont {B.~V.}\ \bibnamefont {Hall}}, \ and\
  \bibinfo {author} {\bibfnamefont {E.~A.}\ \bibnamefont {Hinds}},\ }\href
  {\doibase 10.1103/PhysRevLett.91.080401} {\bibfield  {journal} {\bibinfo
  {journal} {Phys. Rev. Lett.}\ }\textbf {\bibinfo {volume} {91}},\ \bibinfo
  {pages} {080401} (\bibinfo {year} {2003})}\BibitemShut {NoStop}%
\bibitem [{\citenamefont {Harber}\ \emph {et~al.}(2003)\citenamefont {Harber},
  \citenamefont {McGuirk}, \citenamefont {Obrecht},\ and\ \citenamefont
  {Cornell}}]{Harber2003}%
  \BibitemOpen
  \bibfield  {author} {\bibinfo {author} {\bibfnamefont {D.~M.}\ \bibnamefont
  {Harber}}, \bibinfo {author} {\bibfnamefont {J.~M.}\ \bibnamefont {McGuirk}},
  \bibinfo {author} {\bibfnamefont {J.~M.}\ \bibnamefont {Obrecht}}, \ and\
  \bibinfo {author} {\bibfnamefont {E.~A.}\ \bibnamefont {Cornell}},\ }\href
  {\doibase 10.1023/a:1026084606385} {\bibfield  {journal} {\bibinfo  {journal}
  {J. Low Temp. Phys.}\ }\textbf {\bibinfo {volume} {133}},\ \bibinfo {pages}
  {229} (\bibinfo {year} {2003})}\BibitemShut {NoStop}%
\bibitem [{\citenamefont {Lin}\ \emph {et~al.}(2004)\citenamefont {Lin},
  \citenamefont {Teper}, \citenamefont {Chin},\ and\ \citenamefont
  {Vuleti{\'{c}}}}]{Lin2004}%
  \BibitemOpen
  \bibfield  {author} {\bibinfo {author} {\bibfnamefont {Y.-J.}\ \bibnamefont
  {Lin}}, \bibinfo {author} {\bibfnamefont {I.}~\bibnamefont {Teper}}, \bibinfo
  {author} {\bibfnamefont {C.}~\bibnamefont {Chin}}, \ and\ \bibinfo {author}
  {\bibfnamefont {V.}~\bibnamefont {Vuleti{\'{c}}}},\ }\href {\doibase
  10.1103/PhysRevLett.92.050404} {\bibfield  {journal} {\bibinfo  {journal}
  {Phys. Rev. Lett.}\ }\textbf {\bibinfo {volume} {92}},\ \bibinfo {pages}
  {050404} (\bibinfo {year} {2004})}\BibitemShut {NoStop}%
\bibitem [{\citenamefont {Rekdal}\ \emph {et~al.}(2004)\citenamefont {Rekdal},
  \citenamefont {Scheel}, \citenamefont {Knight},\ and\ \citenamefont
  {Hinds}}]{Rekdal2004}%
  \BibitemOpen
  \bibfield  {author} {\bibinfo {author} {\bibfnamefont {P.~K.}\ \bibnamefont
  {Rekdal}}, \bibinfo {author} {\bibfnamefont {S.}~\bibnamefont {Scheel}},
  \bibinfo {author} {\bibfnamefont {P.~L.}\ \bibnamefont {Knight}}, \ and\
  \bibinfo {author} {\bibfnamefont {E.~A.}\ \bibnamefont {Hinds}},\ }\href
  {\doibase 10.1103/PhysRevA.70.013811} {\bibfield  {journal} {\bibinfo
  {journal} {Phys. Rev. A}\ }\textbf {\bibinfo {volume} {70}},\ \bibinfo
  {pages} {013811} (\bibinfo {year} {2004})}\BibitemShut {NoStop}%
\bibitem [{\citenamefont {Henkel}(2005)}]{Henkel2005}%
  \BibitemOpen
  \bibfield  {author} {\bibinfo {author} {\bibfnamefont {C.}~\bibnamefont
  {Henkel}},\ }\href {\doibase 10.1140/epjd/e2005-00188-3} {\bibfield
  {journal} {\bibinfo  {journal} {Eur. Phys. J. D}\ }\textbf {\bibinfo {volume}
  {35}},\ \bibinfo {pages} {59} (\bibinfo {year} {2005})}\BibitemShut {NoStop}%
\bibitem [{\citenamefont {Emmert}\ \emph {et~al.}(2009)\citenamefont {Emmert},
  \citenamefont {Lupa{\c{s}}cu}, \citenamefont {Nogues}, \citenamefont {Brune},
  \citenamefont {Raimond},\ and\ \citenamefont {Haroche}}]{Emmert2009}%
  \BibitemOpen
  \bibfield  {author} {\bibinfo {author} {\bibfnamefont {A.}~\bibnamefont
  {Emmert}}, \bibinfo {author} {\bibfnamefont {A.}~\bibnamefont
  {Lupa{\c{s}}cu}}, \bibinfo {author} {\bibfnamefont {G.}~\bibnamefont
  {Nogues}}, \bibinfo {author} {\bibfnamefont {M.}~\bibnamefont {Brune}},
  \bibinfo {author} {\bibfnamefont {J.-M.}\ \bibnamefont {Raimond}}, \ and\
  \bibinfo {author} {\bibfnamefont {S.}~\bibnamefont {Haroche}},\ }\href
  {\doibase 10.1140/epjd/e2009-00001-5} {\bibfield  {journal} {\bibinfo
  {journal} {Eur. Phys. J. D}\ }\textbf {\bibinfo {volume} {51}},\ \bibinfo
  {pages} {173} (\bibinfo {year} {2009})}\BibitemShut {NoStop}%
\bibitem [{\citenamefont {Sidles}\ \emph {et~al.}(2003)\citenamefont {Sidles},
  \citenamefont {Garbini}, \citenamefont {Dougherty},\ and\ \citenamefont
  {Chao}}]{Sidles2003}%
  \BibitemOpen
  \bibfield  {author} {\bibinfo {author} {\bibfnamefont {J.~A.}\ \bibnamefont
  {Sidles}}, \bibinfo {author} {\bibfnamefont {J.~L.}\ \bibnamefont {Garbini}},
  \bibinfo {author} {\bibfnamefont {W.~M.}\ \bibnamefont {Dougherty}}, \ and\
  \bibinfo {author} {\bibfnamefont {S.-H.}\ \bibnamefont {Chao}},\ }\href
  {\doibase 10.1109/jproc.2003.811796} {\bibfield  {journal} {\bibinfo
  {journal} {Proc. {IEEE}}\ }\textbf {\bibinfo {volume} {91}},\ \bibinfo
  {pages} {799} (\bibinfo {year} {2003})}\BibitemShut {NoStop}%
\bibitem [{\citenamefont {Lamoreaux}(1999)}]{Lamoreaux1999}%
  \BibitemOpen
  \bibfield  {author} {\bibinfo {author} {\bibfnamefont {S.~K.}\ \bibnamefont
  {Lamoreaux}},\ }\href {\doibase 10.1103/PhysRevA.60.1717} {\bibfield
  {journal} {\bibinfo  {journal} {Phys. Rev. A}\ }\textbf {\bibinfo {volume}
  {60}},\ \bibinfo {pages} {1717} (\bibinfo {year} {1999})}\BibitemShut
  {NoStop}%
\bibitem [{\citenamefont {{C. T. Munger, Jr}.}(2005)}]{Munger2005}%
  \BibitemOpen
  \bibfield  {author} {\bibinfo {author} {\bibnamefont {{C. T. Munger, Jr}.}},\
  }\href {\doibase 10.1103/PhysRevA.72.012506} {\bibfield  {journal} {\bibinfo
  {journal} {Phys. Rev. A}\ }\textbf {\bibinfo {volume} {72}},\ \bibinfo
  {pages} {012506} (\bibinfo {year} {2005})}\BibitemShut {NoStop}%
\bibitem [{\citenamefont {Amini}\ \emph {et~al.}(2007)\citenamefont {Amini},
  \citenamefont {{C. T. Munger, Jr.}},\ and\ \citenamefont
  {Gould}}]{Amini2007}%
  \BibitemOpen
  \bibfield  {author} {\bibinfo {author} {\bibfnamefont {J.~M.}\ \bibnamefont
  {Amini}}, \bibinfo {author} {\bibnamefont {{C. T. Munger, Jr.}}}, \ and\
  \bibinfo {author} {\bibfnamefont {H.}~\bibnamefont {Gould}},\ }\href
  {\doibase 10.1103/PhysRevA.75.063416} {\bibfield  {journal} {\bibinfo
  {journal} {Phys. Rev. A}\ }\textbf {\bibinfo {volume} {75}},\ \bibinfo
  {pages} {063416} (\bibinfo {year} {2007})}\BibitemShut {NoStop}%
\bibitem [{\citenamefont {Rabey}\ \emph {et~al.}(2016)\citenamefont {Rabey},
  \citenamefont {Devlin}, \citenamefont {Hinds},\ and\ \citenamefont
  {Sauer}}]{Rabey2016}%
  \BibitemOpen
  \bibfield  {author} {\bibinfo {author} {\bibfnamefont {I.~M.}\ \bibnamefont
  {Rabey}}, \bibinfo {author} {\bibfnamefont {J.~A.}\ \bibnamefont {Devlin}},
  \bibinfo {author} {\bibfnamefont {E.~A.}\ \bibnamefont {Hinds}}, \ and\
  \bibinfo {author} {\bibfnamefont {B.~E.}\ \bibnamefont {Sauer}},\ }\href
  {\doibase 10.1063/1.4966991} {\bibfield  {journal} {\bibinfo  {journal} {Rev.
  Sci. Instrum.}\ }\textbf {\bibinfo {volume} {87}},\ \bibinfo {pages} {115110}
  (\bibinfo {year} {2016})}\BibitemShut {NoStop}%
\bibitem [{\citenamefont {Abel}\ \emph
  {et~al.}(2019{\natexlab{a}})\citenamefont {Abel} \emph {et~al.}}]{Abel2019a}%
  \BibitemOpen
  \bibfield  {author} {\bibinfo {author} {\bibfnamefont {C.}~\bibnamefont
  {Abel}} \emph {et~al.},\ }\href {\doibase 10.1051/epjconf/201921902002}
  {\bibfield  {journal} {\bibinfo  {journal} {{EPJ} Web Conf.}\ }\textbf
  {\bibinfo {volume} {219}},\ \bibinfo {pages} {02002} (\bibinfo {year}
  {2019}{\natexlab{a}})}\BibitemShut {NoStop}%
\bibitem [{\citenamefont {Ayres}\ \emph {et~al.}(2021)\citenamefont {Ayres}
  \emph {et~al.}}]{Ayres2021}%
  \BibitemOpen
  \bibfield  {author} {\bibinfo {author} {\bibfnamefont {N.~J.}\ \bibnamefont
  {Ayres}} \emph {et~al.},\ }\href {\doibase 10.1140/epjc/s10052-021-09298-z}
  {\bibfield  {journal} {\bibinfo  {journal} {Eur. Phys. J. C}\ }\textbf
  {\bibinfo {volume} {81}},\ \bibinfo {pages} {512} (\bibinfo {year}
  {2021})}\BibitemShut {NoStop}%
\bibitem [{\citenamefont {Lauss}(2014)}]{Lauss2014}%
  \BibitemOpen
  \bibfield  {author} {\bibinfo {author} {\bibfnamefont {B.}~\bibnamefont
  {Lauss}},\ }\href {\doibase 10.1016/j.phpro.2013.12.022} {\bibfield
  {journal} {\bibinfo  {journal} {Phys. Proc.}\ }\textbf {\bibinfo {volume}
  {51}},\ \bibinfo {pages} {98} (\bibinfo {year} {2014})}\BibitemShut {NoStop}%
\bibitem [{\citenamefont {Bison}\ \emph {et~al.}(2020)\citenamefont {Bison},
  \citenamefont {Blau}, \citenamefont {Daum}, \citenamefont {Göltl},
  \citenamefont {Henneck}, \citenamefont {Kirch}, \citenamefont {Lauss},
  \citenamefont {Ries}, \citenamefont {Schmidt-Wellenburg},\ and\ \citenamefont
  {Zsigmond}}]{Bison2020}%
  \BibitemOpen
  \bibfield  {author} {\bibinfo {author} {\bibfnamefont {G.}~\bibnamefont
  {Bison}}, \bibinfo {author} {\bibfnamefont {B.}~\bibnamefont {Blau}},
  \bibinfo {author} {\bibfnamefont {M.}~\bibnamefont {Daum}}, \bibinfo {author}
  {\bibfnamefont {L.}~\bibnamefont {Göltl}}, \bibinfo {author} {\bibfnamefont
  {R.}~\bibnamefont {Henneck}}, \bibinfo {author} {\bibfnamefont
  {K.}~\bibnamefont {Kirch}}, \bibinfo {author} {\bibfnamefont
  {B.}~\bibnamefont {Lauss}}, \bibinfo {author} {\bibfnamefont
  {D.}~\bibnamefont {Ries}}, \bibinfo {author} {\bibfnamefont {P.}~\bibnamefont
  {Schmidt-Wellenburg}}, \ and\ \bibinfo {author} {\bibfnamefont
  {G.}~\bibnamefont {Zsigmond}},\ }\href {\doibase
  10.1140/epja/s10050-020-00027-w} {\bibfield  {journal} {\bibinfo  {journal}
  {Eur. Phys. J. A}\ }\textbf {\bibinfo {volume} {56}},\ \bibinfo {pages} {33}
  (\bibinfo {year} {2020})}\BibitemShut {NoStop}%
\bibitem [{\citenamefont {Bison}\ \emph {et~al.}(2017)\citenamefont {Bison},
  \citenamefont {Daum}, \citenamefont {Kirch}, \citenamefont {Lauss},
  \citenamefont {Ries}, \citenamefont {Schmidt-Wellenburg}, \citenamefont
  {Zsigmond}, \citenamefont {Brenner}, \citenamefont {Geltenbort},
  \citenamefont {Jenke} \emph {et~al.}}]{Bison2017}%
  \BibitemOpen
  \bibfield  {author} {\bibinfo {author} {\bibfnamefont {G.}~\bibnamefont
  {Bison}}, \bibinfo {author} {\bibfnamefont {M.}~\bibnamefont {Daum}},
  \bibinfo {author} {\bibfnamefont {K.}~\bibnamefont {Kirch}}, \bibinfo
  {author} {\bibfnamefont {B.}~\bibnamefont {Lauss}}, \bibinfo {author}
  {\bibfnamefont {D.}~\bibnamefont {Ries}}, \bibinfo {author} {\bibfnamefont
  {P.}~\bibnamefont {Schmidt-Wellenburg}}, \bibinfo {author} {\bibfnamefont
  {G.}~\bibnamefont {Zsigmond}}, \bibinfo {author} {\bibfnamefont
  {T.}~\bibnamefont {Brenner}}, \bibinfo {author} {\bibfnamefont
  {P.}~\bibnamefont {Geltenbort}}, \bibinfo {author} {\bibfnamefont
  {T.}~\bibnamefont {Jenke}},  \emph {et~al.},\ }\href {\doibase
  10.1103/PhysRevC.95.045503} {\bibfield  {journal} {\bibinfo  {journal} {Phys.
  Rev. C}\ }\textbf {\bibinfo {volume} {95}},\ \bibinfo {pages} {045503}
  (\bibinfo {year} {2017})}\BibitemShut {NoStop}%
\bibitem [{\citenamefont {Ito}\ \emph {et~al.}(2018)\citenamefont {Ito} \emph
  {et~al.}}]{Ito2018}%
  \BibitemOpen
  \bibfield  {author} {\bibinfo {author} {\bibfnamefont {T.~M.}\ \bibnamefont
  {Ito}} \emph {et~al.},\ }\href {\doibase 10.1103/physrevc.97.012501}
  {\bibfield  {journal} {\bibinfo  {journal} {Phys. Rev. C}\ }\textbf {\bibinfo
  {volume} {97}},\ \bibinfo {pages} {012501(R)} (\bibinfo {year}
  {2018})}\BibitemShut {NoStop}%
\bibitem [{\citenamefont {Atchison}\ \emph {et~al.}(2005)\citenamefont
  {Atchison} \emph {et~al.}}]{Atchison2005}%
  \BibitemOpen
  \bibfield  {author} {\bibinfo {author} {\bibfnamefont {F.}~\bibnamefont
  {Atchison}} \emph {et~al.},\ }\href {\doibase 10.1016/j.physletb.2005.08.066}
  {\bibfield  {journal} {\bibinfo  {journal} {Phys. Lett. B}\ }\textbf
  {\bibinfo {volume} {625}},\ \bibinfo {pages} {19} (\bibinfo {year}
  {2005})}\BibitemShut {NoStop}%
\bibitem [{\citenamefont {Atchison}\ \emph {et~al.}(2006)\citenamefont
  {Atchison} \emph {et~al.}}]{Atchison2006}%
  \BibitemOpen
  \bibfield  {author} {\bibinfo {author} {\bibfnamefont {F.}~\bibnamefont
  {Atchison}} \emph {et~al.},\ }\href {\doibase 10.1103/PhysRevC.74.055501}
  {\bibfield  {journal} {\bibinfo  {journal} {Phys. Rev. C}\ }\textbf {\bibinfo
  {volume} {74}},\ \bibinfo {pages} {055501} (\bibinfo {year}
  {2006})}\BibitemShut {NoStop}%
\bibitem [{\citenamefont {Weis}\ and\ \citenamefont
  {Wynands}(2005)}]{Weis2005}%
  \BibitemOpen
  \bibfield  {author} {\bibinfo {author} {\bibfnamefont {A.}~\bibnamefont
  {Weis}}\ and\ \bibinfo {author} {\bibfnamefont {R.}~\bibnamefont {Wynands}},\
  }\href {\doibase 10.1016/j.optlaseng.2004.03.010} {\bibfield  {journal}
  {\bibinfo  {journal} {Opt. Lasers Eng.}\ }\textbf {\bibinfo {volume} {43}},\
  \bibinfo {pages} {387} (\bibinfo {year} {2005})}\BibitemShut {NoStop}%
\bibitem [{\citenamefont {Groeger}\ \emph {et~al.}(2006)\citenamefont
  {Groeger}, \citenamefont {Bison}, \citenamefont {Schenker}, \citenamefont
  {Wynands},\ and\ \citenamefont {Weis}}]{Groeger2006}%
  \BibitemOpen
  \bibfield  {author} {\bibinfo {author} {\bibfnamefont {S.}~\bibnamefont
  {Groeger}}, \bibinfo {author} {\bibfnamefont {G.}~\bibnamefont {Bison}},
  \bibinfo {author} {\bibfnamefont {J.-L.}\ \bibnamefont {Schenker}}, \bibinfo
  {author} {\bibfnamefont {R.}~\bibnamefont {Wynands}}, \ and\ \bibinfo
  {author} {\bibfnamefont {A.}~\bibnamefont {Weis}},\ }\href {\doibase
  10.1140/epjd/e2006-00037-y} {\bibfield  {journal} {\bibinfo  {journal} {Eur.
  Phys. J. D}\ }\textbf {\bibinfo {volume} {38}},\ \bibinfo {pages} {239}
  (\bibinfo {year} {2006})}\BibitemShut {NoStop}%
\bibitem [{\citenamefont {Green}\ \emph {et~al.}(1998)\citenamefont {Green},
  \citenamefont {Harris}, \citenamefont {Iaydjiev}, \citenamefont {May},
  \citenamefont {Pendlebury}, \citenamefont {Smith}, \citenamefont {van~der
  Grinten}, \citenamefont {Geltenbort},\ and\ \citenamefont
  {Ivanov}}]{Green1998}%
  \BibitemOpen
  \bibfield  {author} {\bibinfo {author} {\bibfnamefont {K.}~\bibnamefont
  {Green}}, \bibinfo {author} {\bibfnamefont {P.~G.}\ \bibnamefont {Harris}},
  \bibinfo {author} {\bibfnamefont {P.}~\bibnamefont {Iaydjiev}}, \bibinfo
  {author} {\bibfnamefont {D.~J.~R.}\ \bibnamefont {May}}, \bibinfo {author}
  {\bibfnamefont {J.~M.}\ \bibnamefont {Pendlebury}}, \bibinfo {author}
  {\bibfnamefont {K.~F.}\ \bibnamefont {Smith}}, \bibinfo {author}
  {\bibfnamefont {M.}~\bibnamefont {van~der Grinten}}, \bibinfo {author}
  {\bibfnamefont {P.}~\bibnamefont {Geltenbort}}, \ and\ \bibinfo {author}
  {\bibfnamefont {S.}~\bibnamefont {Ivanov}},\ }\href {\doibase
  10.1016/s0168-9002(97)01121-2} {\bibfield  {journal} {\bibinfo  {journal}
  {Nucl. Instrum. Methods Phys. Res. A}\ }\textbf {\bibinfo {volume} {404}},\
  \bibinfo {pages} {381} (\bibinfo {year} {1998})}\BibitemShut {NoStop}%
\bibitem [{\citenamefont {Baker}\ \emph {et~al.}(2014)\citenamefont {Baker}
  \emph {et~al.}}]{Baker2014}%
  \BibitemOpen
  \bibfield  {author} {\bibinfo {author} {\bibfnamefont {C.~A.}\ \bibnamefont
  {Baker}} \emph {et~al.},\ }\href {\doibase 10.1016/j.nima.2013.10.005}
  {\bibfield  {journal} {\bibinfo  {journal} {Nucl. Instrum. Methods Phys. Res.
  A}\ }\textbf {\bibinfo {volume} {736}},\ \bibinfo {pages} {184} (\bibinfo
  {year} {2014})}\BibitemShut {NoStop}%
\bibitem [{\citenamefont {Ban}\ \emph {et~al.}(2018)\citenamefont {Ban} \emph
  {et~al.}}]{Ban2018}%
  \BibitemOpen
  \bibfield  {author} {\bibinfo {author} {\bibfnamefont {G.}~\bibnamefont
  {Ban}} \emph {et~al.},\ }\href {\doibase 10.1016/j.nima.2018.04.025}
  {\bibfield  {journal} {\bibinfo  {journal} {Nucl. Instrum. Methods Phys. Res.
  A}\ }\textbf {\bibinfo {volume} {896}},\ \bibinfo {pages} {129} (\bibinfo
  {year} {2018})}\BibitemShut {NoStop}%
\bibitem [{Note1()}]{Note1}%
  \BibitemOpen
  \bibinfo {note} {Compare to Eq.\protect \tmspace +\thinmuskip {.1667em}(38)
  of Ref.\protect \tmspace +\thinmuskip {.1667em}\cite {Varpula1984}, where
  variables $z$, distance, and $t$, thickness, were replaced with $d$ and $a$
  in our paper to avoid confusion with other variables used in the remainder of
  the article.}\BibitemShut {Stop}%
\bibitem [{\citenamefont {Bork}\ \emph {et~al.}(2001)\citenamefont {Bork},
  \citenamefont {Hahlbohm}, \citenamefont {Klein},\ and\ \citenamefont
  {Schnabel}}]{Bork2001}%
  \BibitemOpen
  \bibfield  {author} {\bibinfo {author} {\bibfnamefont {J.}~\bibnamefont
  {Bork}}, \bibinfo {author} {\bibfnamefont {H.-D.}\ \bibnamefont {Hahlbohm}},
  \bibinfo {author} {\bibfnamefont {R.}~\bibnamefont {Klein}}, \ and\ \bibinfo
  {author} {\bibfnamefont {A.}~\bibnamefont {Schnabel}},\ }in\ \href
  {https://research.aalto.fi/en/publications/biomag-2000-proceedings-of-the-12th-international-conference-on-b}
  {\emph {\bibinfo {booktitle} {Biomag2000, Proc. 12$^{th}$ Int. Conf. on
  Biomagnetism}}}\ (\bibinfo {year} {2001})\ pp.\ \bibinfo {pages}
  {970--973}\BibitemShut {NoStop}%
\bibitem [{\citenamefont {Thiel}\ \emph {et~al.}(2005)\citenamefont {Thiel},
  \citenamefont {Schnabel}, \citenamefont {Knappe-Gr\"{u}neberg}, \citenamefont
  {Burghoff}, \citenamefont {Drung}, \citenamefont {Petsche}, \citenamefont
  {Bechstein}, \citenamefont {Steinhoff}, \citenamefont {M\"{u}ller},
  \citenamefont {Stollfu{\ss}} \emph {et~al.}}]{Thiel2005}%
  \BibitemOpen
  \bibfield  {author} {\bibinfo {author} {\bibfnamefont {F.}~\bibnamefont
  {Thiel}}, \bibinfo {author} {\bibfnamefont {A.}~\bibnamefont {Schnabel}},
  \bibinfo {author} {\bibfnamefont {S.}~\bibnamefont {Knappe-Gr\"{u}neberg}},
  \bibinfo {author} {\bibfnamefont {M.}~\bibnamefont {Burghoff}}, \bibinfo
  {author} {\bibfnamefont {D.}~\bibnamefont {Drung}}, \bibinfo {author}
  {\bibfnamefont {F.}~\bibnamefont {Petsche}}, \bibinfo {author} {\bibfnamefont
  {S.}~\bibnamefont {Bechstein}}, \bibinfo {author} {\bibfnamefont
  {U.}~\bibnamefont {Steinhoff}}, \bibinfo {author} {\bibfnamefont
  {W.}~\bibnamefont {M\"{u}ller}}, \bibinfo {author} {\bibfnamefont
  {D.}~\bibnamefont {Stollfu{\ss}}},  \emph {et~al.},\ }\href
  {https://www.researchgate.net/profile/F_Thiel/publication/222396471_The_304_SQUIDs_vector_magnetometer_system_for_biomagnetic_measurements_in_the_Berlin_Magnetically_Shielded_Room_2/links/0912f4ffec79fa4728000000/The-304-SQUIDs-vector-magnetometer-system-for-biomagnetic-measurements-in-the-Berlin-Magnetically-Shielded-Room-2.pdf}
  {\bibfield  {journal} {\bibinfo  {journal} {Biomed. Technik}\ }\textbf
  {\bibinfo {volume} {50}},\ \bibinfo {pages} {169} (\bibinfo {year}
  {2005})}\BibitemShut {NoStop}%
\bibitem [{\citenamefont {Abel}\ \emph
  {et~al.}(2019{\natexlab{b}})\citenamefont {Abel} \emph {et~al.}}]{Abel2019}%
  \BibitemOpen
  \bibfield  {author} {\bibinfo {author} {\bibfnamefont {C.}~\bibnamefont
  {Abel}} \emph {et~al.},\ }\href {\doibase 10.1103/PhysRevA.99.042112}
  {\bibfield  {journal} {\bibinfo  {journal} {Phys. Rev. A}\ }\textbf {\bibinfo
  {volume} {99}},\ \bibinfo {pages} {042112} (\bibinfo {year}
  {2019}{\natexlab{b}})}\BibitemShut {NoStop}%
\bibitem [{\citenamefont {Caflisch}(1998)}]{Caflisch1998}%
  \BibitemOpen
  \bibfield  {author} {\bibinfo {author} {\bibfnamefont {R.~E.}\ \bibnamefont
  {Caflisch}},\ }\href {\doibase 10.1017/s0962492900002804} {\bibfield
  {journal} {\bibinfo  {journal} {Acta Numerica}\ }\textbf {\bibinfo {volume}
  {7}},\ \bibinfo {pages} {1} (\bibinfo {year} {1998})}\BibitemShut {NoStop}%
\bibitem [{\citenamefont {Lee}\ and\ \citenamefont {Romalis}(2008)}]{Lee2008}%
  \BibitemOpen
  \bibfield  {author} {\bibinfo {author} {\bibfnamefont {S.-K.}\ \bibnamefont
  {Lee}}\ and\ \bibinfo {author} {\bibfnamefont {M.~V.}\ \bibnamefont
  {Romalis}},\ }\href {\doibase 10.1063/1.2885711} {\bibfield  {journal}
  {\bibinfo  {journal} {J. Appl. Phys.}\ }\textbf {\bibinfo {volume} {103}},\
  \bibinfo {pages} {084904} (\bibinfo {year} {2008})}\BibitemShut {NoStop}%
\bibitem [{Note2()}]{Note2}%
  \BibitemOpen
  \bibinfo {note} {Compare to the original expression shown in Ref.\protect
  \tmspace +\thinmuskip {.1667em}\cite {Lee2008}, where variables $a$,
  distance, $t$, thickness, and $r$, radius, were replaced with $d$, $a$ and
  $R$ in our paper to avoid confusion and keep consistency with other variables
  used in the remainder of the article.}\BibitemShut {Stop}%
\bibitem [{\citenamefont {Zsigmond}(2018)}]{Zsigmond2018}%
  \BibitemOpen
  \bibfield  {author} {\bibinfo {author} {\bibfnamefont {G.}~\bibnamefont
  {Zsigmond}},\ }\href {\doibase 10.1016/j.nima.2017.10.065} {\bibfield
  {journal} {\bibinfo  {journal} {Nucl. Instrum. Methods Phys. Res. A}\
  }\textbf {\bibinfo {volume} {881}},\ \bibinfo {pages} {16} (\bibinfo {year}
  {2018})}\BibitemShut {NoStop}%
\bibitem [{\citenamefont {Abel}\ \emph {et~al.}(2020)\citenamefont {Abel} \emph
  {et~al.}}]{Abel2020}%
  \BibitemOpen
  \bibfield  {author} {\bibinfo {author} {\bibfnamefont {C.}~\bibnamefont
  {Abel}} \emph {et~al.},\ }\href {\doibase 10.1103/PhysRevLett.124.081803}
  {\bibfield  {journal} {\bibinfo  {journal} {Phys. Rev. Lett.}\ }\textbf
  {\bibinfo {volume} {124}},\ \bibinfo {pages} {081803} (\bibinfo {year}
  {2020})}\BibitemShut {NoStop}%
\bibitem [{\citenamefont {Bell}\ and\ \citenamefont {Bloom}(1961)}]{Bell1961}%
  \BibitemOpen
  \bibfield  {author} {\bibinfo {author} {\bibfnamefont {W.~E.}\ \bibnamefont
  {Bell}}\ and\ \bibinfo {author} {\bibfnamefont {A.~L.}\ \bibnamefont
  {Bloom}},\ }\href {\doibase 10.1103/PhysRevLett.6.280} {\bibfield  {journal}
  {\bibinfo  {journal} {Phys. Rev. Lett.}\ }\textbf {\bibinfo {volume} {6}},\
  \bibinfo {pages} {280} (\bibinfo {year} {1961})}\BibitemShut {NoStop}%
\bibitem [{\citenamefont {Gruji{\'{c}}}\ \emph {et~al.}(2015)\citenamefont
  {Gruji{\'{c}}}, \citenamefont {Koss}, \citenamefont {Bison},\ and\
  \citenamefont {Weis}}]{Grujic2015}%
  \BibitemOpen
  \bibfield  {author} {\bibinfo {author} {\bibfnamefont {Z.~D.}\ \bibnamefont
  {Gruji{\'{c}}}}, \bibinfo {author} {\bibfnamefont {P.~A.}\ \bibnamefont
  {Koss}}, \bibinfo {author} {\bibfnamefont {G.}~\bibnamefont {Bison}}, \ and\
  \bibinfo {author} {\bibfnamefont {A.}~\bibnamefont {Weis}},\ }\href {\doibase
  10.1140/epjd/e2015-50875-3} {\bibfield  {journal} {\bibinfo  {journal} {Eur.
  Phys. J. D}\ }\textbf {\bibinfo {volume} {69}},\ \bibinfo {pages} {135}
  (\bibinfo {year} {2015})}\BibitemShut {NoStop}%
\bibitem [{\citenamefont {Afach}\ \emph {et~al.}(2015)\citenamefont {Afach}
  \emph {et~al.}}]{Afach2015d}%
  \BibitemOpen
  \bibfield  {author} {\bibinfo {author} {\bibfnamefont {S.}~\bibnamefont
  {Afach}} \emph {et~al.},\ }\href {\doibase 10.1364/oe.23.022108} {\bibfield
  {journal} {\bibinfo  {journal} {Opt. Express}\ }\textbf {\bibinfo {volume}
  {23}},\ \bibinfo {pages} {22108} (\bibinfo {year} {2015})}\BibitemShut
  {NoStop}%
\bibitem [{\citenamefont {Griffiths}(2017)}]{Griffiths2017}%
  \BibitemOpen
  \bibfield  {author} {\bibinfo {author} {\bibfnamefont {D.~J.}\ \bibnamefont
  {Griffiths}},\ }\href {\doibase 10.1017/9781108333511} {\emph {\bibinfo
  {title} {Introduction to Electrodynamics}}},\ \bibinfo {edition} {4th}\ ed.\
  (\bibinfo  {publisher} {Cambridge University Press},\ \bibinfo {year}
  {2017})\BibitemShut {NoStop}%
\bibitem [{\citenamefont {Jackson}(1998)}]{Jackson1998}%
  \BibitemOpen
  \bibfield  {author} {\bibinfo {author} {\bibfnamefont {J.~D.}\ \bibnamefont
  {Jackson}},\ }\href {https://www.wiley.com/en-us/Classical Electrodynamics,
  3rd Edition-p-9780471309321} {\emph {\bibinfo {title} {Classical
  Electrodynamics}}},\ \bibinfo {edition} {3rd}\ ed.\ (\bibinfo  {publisher}
  {WILEY},\ \bibinfo {year} {1998})\BibitemShut {NoStop}%
\end{thebibliography}%
\end{document}